\documentclass{SciPost}

\usepackage{graphicx}
\usepackage{hyperref}
\usepackage{mathrsfs}
\usepackage{amsmath}
\usepackage{xcolor}
\usepackage{siunitx}

\hypersetup{
    colorlinks,
    linkcolor={red!50!black},
    citecolor={blue!50!black},
    urlcolor={blue!80!black}
}

\usepackage[bitstream-charter]{mathdesign}
\DeclareSymbolFont{usualmathcal}{OMS}{cmsy}{m}{n}
\DeclareSymbolFontAlphabet{\mathcal}{usualmathcal}

\fancypagestyle{SPstyle}{
\fancyhf{}
\lhead{\colorbox{scipostblue}{\bf \color{white} ~SciPost Physics }}
\rhead{{\bf \color{scipostdeepblue} ~Submission }}

\fancyfoot[C]{\textbf{\thepage}}
}

\begin{document}

\pagestyle{SPstyle}

\begin{center}{\Large \textbf{\color{scipostdeepblue}{
%%%%%%%%%% TODO: Write your article's title here
Current-induced re-entrant superconductivity and extreme nonreciprocal superconducting diode effect in valley-polarized systems\\
%%%%%%%%%% END TODO: TITLE
}}}\end{center}

\begin{center}\textbf{
%%%%%%%%%% TODO: AUTHORS
% Write the author list here.
% Use (full) first name (+ middle name initials) + surname format.
% Separate subsequent authors by a comma, omit comma and use "and" for the last author.
% Mark the corresponding author(s) with a superscript symbol in this order
% \star, \dagger, \ddagger, \circ, \S, \P, \parallel, ...
Yu-Chen Zhuang\textsuperscript{1} and
Qing-Feng Sun\textsuperscript{1, 2$\star$}
%%%%%%%%%% END TODO: AUTHORS
}\end{center}

\begin{center}
%%%%%%%%%% TODO: AFFILIATIONS
% Write all affiliations here.
% Format: institute, city, country
{\bf 1} International Center for Quantum Materials, School of Physics, Peking University, 100871 Beijing, China
\\
{\bf 2} Hefei National Laboratory, Hefei, 230088 Anhui, China
%%%%%%%%%% END TODO: AFFILIATIONS
%%%%%%%%%% TODO: EMAIL
% Provide email address of corresponding author(s)
\\[\baselineskip]
$\star$ \href{sunqf@pku.edu.cn}{\small sunqf@pku.edu.cn}
%\,,\quad
%$\dagger$ \href{mailto:email2}{\small email2}
%%%%%%%%%% END TODO: EMAIL
\end{center}

\section*{\color{scipostdeepblue}{Abstract}}
%\boldmath
\textbf{%
%%%%%%%%%% TODO: ABSTRACT
% Write your abstract here.
The superconducting diode effect (SDE) refers to the nonreciprocity of superconducting critical currents. Generally, the SDE has a positive and a negative critical currents $j_{c\pm}$ corresponding to two opposite directions with unequal amplitudes. It is demonstrated that an extreme nonreciprocity
where two critical currents can become both positive (or negative) has been observed in twisted graphene systems. In this work, we theoretically propose a possible mechanism to realize an extreme nonreciprocal SDE. Based on a simple microscopic model, we demonstrate that depairing currents required to dissolve Cooper pairs can be remodulated under the interplay between valley polarizations
and applied currents. Near the disappearance of the superconductivity, the remodulation is shown to induce extreme nonreciprocity and also the current-induced re-entrant superconductivity where the system has two different critical current intervals. Our study may provide new horizons for understanding the coexistence of superconductivity and spontaneous valley polarizations, and pave a way for designing SDE with 100\% efficiency.
%%%%%%%%%% END TODO: ABSTRACT
}

\vspace{\baselineskip}

%%%%%%%%%% BLOCK: Copyright information
% This block will be filled during the proof stage, and finilized just before publication.
% It exists here only as a placeholder, and should not be modified by authors.
\noindent\textcolor{white!90!black}{%
\fbox{\parbox{0.975\linewidth}{%
\textcolor{white!40!black}{\begin{tabular}{lr}%
  \begin{minipage}{0.6\textwidth}%
    {\small Copyright attribution to authors. \newline
    This work is a submission to SciPost Physics. \newline
    License information to appear upon publication. \newline
    Publication information to appear upon publication.}
  \end{minipage} & \begin{minipage}{0.4\textwidth}
    {\small Received Date \newline Accepted Date \newline Published Date}%
  \end{minipage}
\end{tabular}}
}}
}
%%%%%%%%%% BLOCK: Copyright information

%%%%%%%%%% TODO: LINENO
% For convenience during refereeing we turn on line numbers:
%\linenumbers
% You should run LaTeX twice in order for the line numbers to appear.
%%%%%%%%%% END TODO: LINENO

%%%%%%%%%% TODO: TOC
% Guideline: if your paper is longer that 6 pages, include a TOC
% To remove the TOC, simply cut the following block
\vspace{10pt}
\noindent\rule{\textwidth}{1pt}
\tableofcontents
\noindent\rule{\textwidth}{1pt}
\vspace{10pt}
%%%%%%%%%% END TODO: TOC

%%%%%%%%% TODO: CONTENTS
% Write your article contents here, starting from first \section.
% An example structure is given below.

\section{\label{SEC1} Introduction}

% TODO: write your article here.
Superconducting diode effect (SDE) is a recently observed superconducting phenomenon with a nonreciprocity of the non-dissipative supercurrents \cite{Nadeem,Jiang}, and has been attracting substantial attention. Such nonreciprocity means amplitudes of critical currents required to destroy the superconductivity are unequal in opposite directions.
As a novel transport phenomenon, SDE can not only uncover
underlying features in exotic superconducting systems \cite{AlfonsoSciPost2024, LuPRB2025},
but also serve as a non-dissipative circuit which has promising applications in low-power superconducting
electronics \cite{Braginski}, superconducting spintronics \cite{Linder, MaoPRL2024}, quantum information and communication technology \cite{Wendin, LiuX}. Since the observation of SDE in artificial superlattice $\mathrm{[Nb/V/Ta]_{n}}$ \cite{Ando},
similar nonreciprocity of supercurrents has been observed
in series of experiments, including bulk materials of
diverse dimensions \cite{Miyasaka,Itahashi,Narita,Masuko,Lin,Bauriedl}, Josephson junction devices \cite{Baumgartner,Jeon,Wu,Pal,deVries,Diez,Turini,Trahms}, engineered superconducting structures \cite{Lyu,Strambini}.
In theory, the rise of SDE usually relies on simultaneous
breaking of time-reversal symmetry (TRS) and inversion symmetry \cite{JorgeSciPost2024,YerinPRB,ChengQ},
which is closely related to magnetochiral anisotropy \cite{Rikken,Rikken2,Wakatsuki,Wakatsuki2,He}, and finite-momentum Cooper pairings \cite{Barzykin,Michael,Yuan,Scammell}.

The performance of the SDE can be measured by the
superconducting diode efficiency
$\eta=\frac{j_{c+}-|j_{c-}|}{j_{c+}+| j_{c-}\vert}$,
with the critical currents $j_{c\pm}$ for positive and
negative directions \cite{SunPRB2025, SunPRB2025-2}. The value of $\eta$ generally depends on the relevant system parameters like working temperature, applied magnetic field and chemical potentials \cite{Daido,Yuan2,He,Ilic}. In most experiments, $\eta$ can be optimized to several tens of percent. One notable exception appears in the experiment for zero-field SDE in small-twist-angle trilayer graphene where critical currents $j_{c\pm}$ are found to even cross zero and become both positive or negative at some regimes\cite{Lin}. This so-called extreme nonreciprocity indicates a realization of SDE with 100\% efficiency. It is a very counterintuitive feature since the electric current does not destroy superconductivity as traditionally believed, but rather promotes a normal state into a superconducting state. Some recent theoretical studies implies that the coupling between the symmetry-breaking order parameter and supercurrents could significantly enhance SDE efficiency $\eta$ \cite{Banerjee}, and dissipations induced by the out-of plane electric field may falicite $100\%$ SDE efficiency \cite{Daido2}. However, the emergence of extreme nonreciprocal SDE still remains an issue that requires further illuminations.

Due to its unique massless Dirac dispersion, the graphene system has been an excellent platform for exploring various novel physical phenomena \cite{KimNature2022, ZhengNC2022, MaoNature2025}. Except for the charge and spin, the electrons in graphene have an additional degree of freedom, valley \cite{JuanSciPost2025, RenPRL2022}. In twisted graphene systems, a non-negligible phenomenon is that a dc current can modulate and even switch the valley polarization \cite{Sharpe,Serlin,Su,Ying,HeWY,Huang}. From the view of the bulk transport, the applied current can redistribute electron occupations in different valley bands near the Fermi level, and then induce energy band shifts due to the Coulomb interaction \cite{Su}. Considering the spontaneous valley polarization plays an important role in the SDE in twisted trilayer graphene, it is worth investigating the connection between the extreme nonreciprocal SDE and the current-induced valley polarization modulation.

In this work, based on the current-induced valley polarization modulation, we theoretically propose a possible mechanism to achieve the extreme nonreciprocity. By a simple valley-polarized system with intervalley pairings, we first study the nonreciprocity of intrinsic depairing current $\tilde{j}_{c}$ \cite{Daido}. Then, we point out $\tilde{j}_{c}$ should be further remodulated to the actual critical current $j_{c}$ because of the interplay between the current and valley occupations. In a large valley splitting regime close to the disappearance of superconductivity, this remodulation could lead to extreme nonreciprocity. The effects of variations of fillings and external magnetic fields on $j_{c}$ are also investigated. Moreover, we raise a new phenomenon, the current-induced re-entrant superconductivity, where the system has two different superconducting regions with distinct critical current intervals. Our study provides a possible routine to achieve SDE with 100\% efficiency and also sheds light on the extreme nonreciprocity observed in the twisted graphene experiment.

The remainder of this article is organized as follows. In Sec.~\ref{SEC2}, we demonstrate our theoretical formalism to achieve the extreme nonreciprocity. In Sec.~\ref{S21}, we first construct a microscopic model to describe the spontaneous valley polarization. In Sec.~\ref{S22}, based on a simple valley-polarized model, we further show a physical mechanism to illustrate the current-induced valley polarization modulation. In Sec.~\ref{S23}, we study superconducting depairing currents and demonstrate a physical process to show how intrinsic depairing currents are remodulated as actual critical currents measured in the experiment. In Sec.~\ref{SEC3}, with a specific model, we use numerical calculations to verify our proposed physical mechanism. The critical currents before and after the remodulation, the variation of actual critical currents with electron occupations and external magnetic fields are investigated in detail, respectively. In Sec.~\ref{SEC4}, we give some discussions and a brief conclusion. The detailed formulations of the current-induced valley polarization modulation are shown in Appendix.~\ref{SEC-A}. In Appendix.~\ref{SEC-B}, we give some theoretical discussions to evaluate the self-consistent manner due to the effect of applied currents. In Appendix.~\ref{SEC-C} and \ref{SEC-D}, we give some discussions about the trigonal warping effect and the coupling between supercurrents and valley polarizations. An estimation of the modulation coefficient $\alpha_{\pm}$ of currents and the effect of band asymmetry are shown in Appendix.~\ref{SEC-E}. The demonstration for the convergence of our results for the system size is put in Appendix.~\ref{SEC-F}.

\section{\label{SEC2} Formalism}

\begin{figure}[ht]
	\includegraphics[width=0.77\columnwidth]{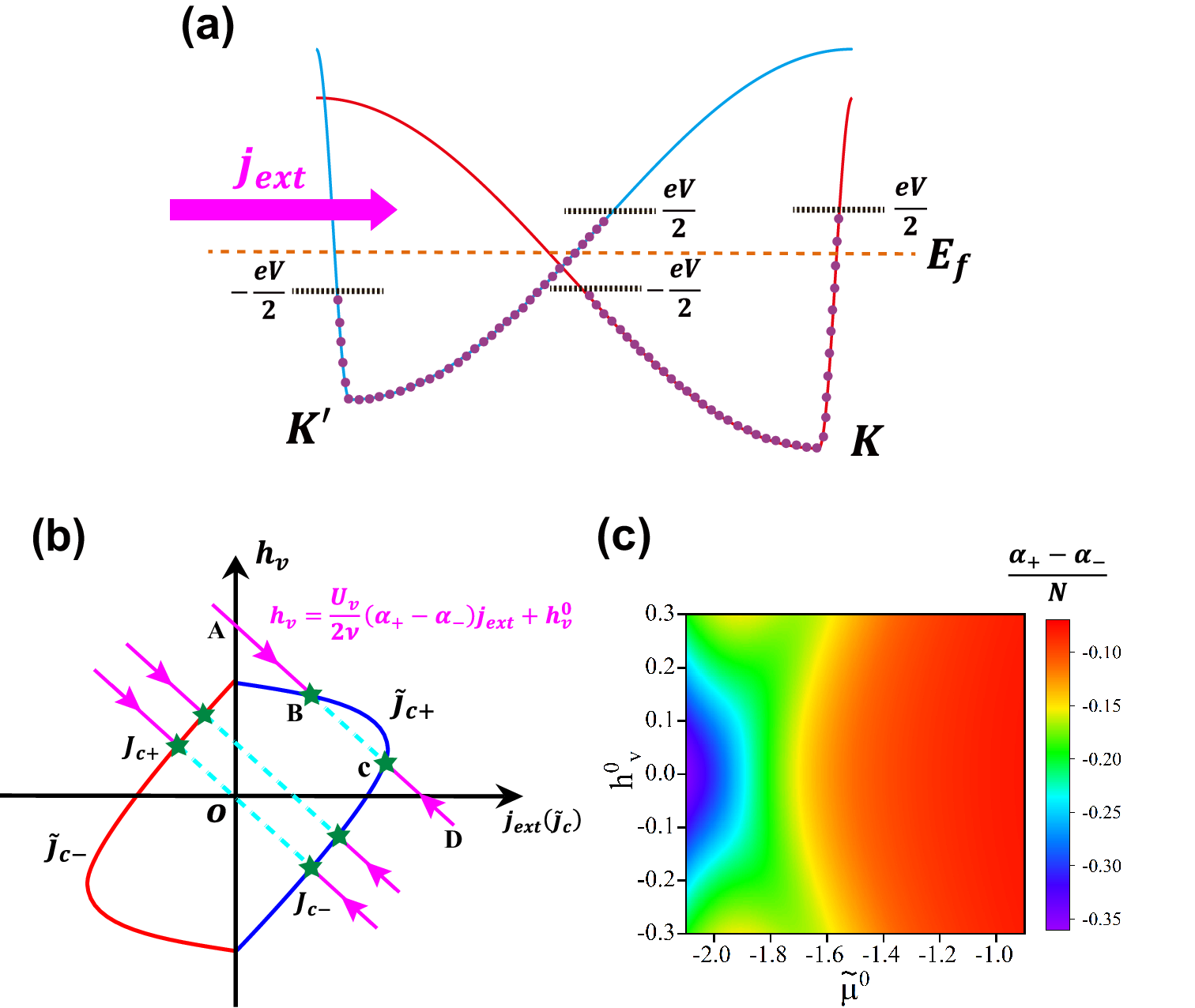}
	\centering %
	\caption{(a) Schematic illustration of the current-induced valley polarization modulation. The red and blue solid line denote two effective valley bands with the index $\tau=\pm$ (i.e., $K$, $K'$). The purple dots denote the occupied electrons on each band. Once an electric current $j_{ext}$ (magenta arrow line) is applied, the local Fermi level for electrons with positive (negative) group velocities will climb (descend) by $eV/2$ relative to Fermi level $E_f$ in equilibrium (black dashed lines). (b) The red and dark blue solid lines respectively denote intrinsic depairing currents $\tilde{j}_{c\pm}$ versus the valley splitting field $h_{v}$. The magenta solid lines and cyan dashed lines denote several $h_v-j_{ext}$ relations of Eq.~(\ref{Eq5}) with different $h^0_{v}$. At magenta solid lines, the system stays in the normal phase. While for cyan dashed lines, the system has entered the superconducting phase. The magenta arrows denote the direction of the phase transition starting from the normal phase to the superconducting phase, which is the focus of our theory. The intersection points (dark green stars) denote predicted actual critical currents $j_{c\pm}$. (c) The colormap for $(\alpha_{+}-\alpha_{-})/N$ versus $h^0_{v}$ and $\tilde{\mu}^0$.}
	\label{FIG1}
\end{figure}

\subsection{\label{S21} The mean-field model and the interaction-induced valley polarization}

To depict the spontaneous valley polarizations, We simply consider a two-band Hamiltonian to implement a valley-polarized system with an intervalley interaction \cite{Su}:
\begin{equation}
	H^{v}=\sum_{k,\tau}(\epsilon_{k,\tau}-\mu)c_{k,\tau}^{\dagger}c_{k,\tau}
+\frac{U_{v}}{\mathcal{V}}\sum_{k,k^{'}}c^{\dagger}_{k,+}c_{k,+}
c^{\dagger}_{k^{'},-}c_{k^{'},-},
	\label{Eq1}
\end{equation}
where $\tau=\pm$ label the valley index $K,K'$,
$U_{v}>0$ denotes the repulsive intervalley interaction. $\mathcal{V}$ and $\mu$ are the systemic size and chemical potential, respectively.
$\epsilon_{k,\tau}$ denotes the single-particle band, which satisfies TRS: $\epsilon_{k,+}=\epsilon_{-k,-}$. Taking the mean-field approximation, the Hamiltonian becomes $H_{MF}^{v}=\sum_{k,\tau}E_{k,\tau}c^{\dagger}_{k,\tau}c_{k,\tau}+$ const where $E_{k,\tau}=\epsilon_{k,\tau}-\mu+\frac{U_v}{\mathcal{V}}n_{-\tau}$. Here $n_{\tau}$ denotes the average electron occupation for $\tau$ valley:  $n_{\tau}=\sum_{k}\langle c^{\dagger}_{k,\tau}c_{k,\tau}\rangle=\sum_{k}f(E_{k,\tau})$ with the Fermi distribution $f(E_{k,\tau})=1/(1+e^\frac{E_{k,\tau}}{T})$ ($T$ is the thermal energy). The $\mathrm{const}=-\frac{U_{v}}{\mathcal{V}}n_{+}n_{-}$ is a constant arising from the mean-field approximation.
Note that this mean-field model is similar to the rigid band flavor Stoner model with a $SU(4)$ symmetric Coulomb interaction energy $V_{int}\propto \sum_{\alpha \neq \beta}n_{\alpha}n_{\beta}$ ($\alpha$,$\beta$ denote four flavors $K\uparrow,K\downarrow,K'\uparrow,K'\downarrow$), which is often used to study flavor polarizations in graphene \cite{Zondiner,Zhou}. Since the valley polarization plays the most important role in the experiment \cite{Lin}, we first focus on valley flavors and neglect spin flavors.

It is easy to find that the growth of the electron occupation $n_{\tau}$ could lift the energy of $-\tau$ valley, and thus influence the free energy $F_{v}$ of the system:
\begin{equation}
	F_{v}(n,m)=-T\sum_{k,\tau}\ln(1+e^{-\frac{E_{k,\tau}}{T}})-\frac{U_{v}}{4\mathcal{V}}(n^{2}-m^{2})+\mu n,
	\label{Eq2}
\end{equation}
where $n=n_{+}+n_{-}$ is the total electron occupation, $m=n_{+}-n_{-}$ denotes the valley polarization. Generally, the system is fixed with a definite total electron occupation $n$ and reaches a state where $m$ is just the minimum point of the free energy $F_{v}(n,m)$. Therefore, the valley polarization can be solved by:
\begin{equation}
	\frac{\partial F_{v}}{\partial m}=\frac{U_{v}}{2\mathcal{V}}
	\left(m-\sum_{k,\tau}\frac{\tau}{1+e^{\frac{E_{k,\tau}}{T}}}\right)=0.
	\label{Eq3}
\end{equation}
Once $U_{v}g(E_{f})/\mathcal{V}>1$ where $g(E_{f})$ is the density of states at the Fermi level $E_{f}$, the strong repulsive Coulomb interaction overwhelms the kinetic energy and make the system favor a nonequal electron distribution between two valleys. This is analogous to the well-known Stoner criterion and at this time the solution $m$ in Eq.~(\ref{Eq3}) is nonzero \cite{Stoner}. The spontaneous valley polarization further introduces a valley splitting field $h_{v}=\frac{U_{v}m}{2\mathcal{V}}$ and a modified chemical potential $\tilde{\mu}=\mu-\frac{U_{v}n}{2\mathcal{V}}$ in mean-field bands $E_{k,\tau}=\epsilon_{k,\tau}-\tilde{\mu}-h_{v}\tau$.

\subsection{\label{S22} The current-induced valley polarization modulations}

In twisted graphene systems, it is found that a dc current could modulate and even switch the valley polarization
\cite{Sharpe,Serlin,Su,Ying,HeWY,Huang}.
We here illustrate it from a nonequilibrium ballistic quantum transport. In Fig.~\ref{FIG1}(a), the red solid line and blue solid line schematically correspond to two bands with valley $K$ and valley $K'$, respectively. Due to intervalley interaction, there is a spontaneous symmetry breaking as a consequence of two valley band splittings (e.g., the red band is below the blue band). Then, under an external bias $V$, an applied current $j_{ext}$ (magenta arrow) will flow through the system and be carried by the energy bands. It leads to the Fermi level of electrons with positive (negative) velocities will rise (fall), for simplicity, $\frac{eV}{2}$ relative to the Fermi level $E_{f}$ in equilibrium \cite{Datta}. Moreover, if the valley bands are intravalley inversion symmetry-broken bands (i.e., $\epsilon_{k,\tau} \neq \epsilon_{-k,\tau}$),
the variation of electron occupation for opposite velocities cannot be offset due to unequal density of states at the Fermi level, as indicated by purple dots (electron occupations) on colored solid lines in Fig.~\ref{FIG1}(a).
Therefore, $n_{\tau}$ will be changed in each valley and is approximately proportional to $j_{ext}$ at a small bias $V$ (see detailed derivations in Appendix.~\ref{SEC-A}):
\begin{equation}
n_{\tau}=n^{0}_{\tau}+\alpha_{\tau}j_{ext}.
	\label{Eq4}
\end{equation}
The modulation coefficient $\alpha_{\tau}$ is a function of the modified chemical potential $\tilde{\mu}$ and also the valley splitting field $h_{v}$. It relies on the difference between the positive and negative Fermi velocities and will be zero once $\epsilon_{k,\tau} = \epsilon_{-k,\tau}$ (see Appendix.~\ref{SEC-A}), which is again consistent with our picture shown in Fig.~\ref{FIG1}(a). 

The variation of $n_{\tau}$ in Eq.~(\ref{Eq4}) will further alter valley polarization $m=n_{+}-n_{-}$ and the valley splitting field:
\begin{equation}	
	h_{v}=\frac{U_{v}}{2\mathcal{V}}m=\frac{U_{v}}{2\mathcal{V}}(\alpha_{+}-\alpha_{-})j_{ext}+h_{v}^{0}.
	\label{Eq5}
\end{equation}
Here $h_{v}^0$ are the initial valley splitting field at $j_{ext} = 0$. It should be noted that the linear relation in Eq.~(\ref{Eq5}) is only an approximation. In principle, the applied current $j_{ext}$ which redistributes electron occupations in each valley can also refresh the value of $h_{v}^{0}$ simultaneously. The rigorous self-consistent calculation of $h_{v}^{0}$ including the nonequilibrium electric current is a subtle question. For simplicity, we focus on a small current range where the impact of $j_{ext}$ on $h^{0}_{v}$ should be minor (see further discussions in Appendix.~\ref{SEC-B}). And thus in the calculations, we just ignore the influence of $j_{ext}$ on $h_{v}^{0}$ in the right side of Eq.~(\ref{Eq5}). We fix the modified chemical potential $\tilde{\mu}=\tilde{\mu}^0$ (considering the electron occupation is fixed) and ignore the dependence of $\alpha_{\pm}$ on $h_v$ (consider the variation of $h_{v}$ is not large at a small current). Then, the coefficient $\alpha_{\pm}$ on the right of Eq.~(\ref{Eq5}) is just set as $\alpha_{\pm}(h_{v}^0, \tilde{\mu}^0)$, for simplicity.

The breaking of intravalley inversion symmetry on the energy bands could naturally exist in twisted graphene systems \cite{Su},
as well as some materials with trigonal warping effect on the Fermi surface \cite{Hu} (see more discussions in Appendix.~\ref{SEC-C}). Additionally, TRS guarantees opposite signs of $\alpha_{\pm}$. See Fig.~\ref{FIG1}(a), $j_{ext}$ will make $n_{-}$ larger and $n_{+}$ smaller, thereby reducing the valley polarization $m$ and valley splitting field $h_{v}$.

\subsection{\label{S23} The remodulation of critical currents and extreme nonreciprocity}

Based on the valley-polarized system shown in Sec.~\ref{S21}, we further consider an s-wave finite-momentum intervalley pairing $H_{s}=-\frac{U_{s}}{\mathcal{V}} \sum_{k,q}c^{\dagger}_{k+q,+}c^{\dagger}_{-k+q,-}c_{-k+q,-}c_{k+q,+}$ in the system. Here $2q$ denotes the center-of-mass momentum of Cooper pairs. Although there is a competition between valley ferromagnetism and superconductivity, the traits of the coexistence between them have been found in twisted graphene systems \cite{Lin,ZhangNai}. To simplify the problem, we here consider the spontaneous valley polarization and superconducting pairing as two separate steps. At first, we regard the system as a normal state with valley-polarized bands determined from the mean-field solution in Eq.~(\ref{Eq1}). Next, following ref.~\cite{Daido}, we further add the s-wave intervalley pairing on it and now focus on its Bardeen-Cooper-Schrieffer (BCS) mean-field Hamiltonian for each fixed $q$:
\begin{equation}
  H = \sum_{k,\tau}E_{k,\tau}c^{\dagger}_{k,\tau}c_{k,\tau}-
\sum_{k,q}\Delta(q)c^{\dagger}_{k+q,+}c^{\dagger}_{-k+q,-}+\mathrm{H.c}.
\label{Eq6}
\end{equation}
where the first term is from the mean-field Hamiltonian of Eq.~(\ref{Eq1}) and the second term denotes s-wave intervalley superconducting order parameter $\Delta(q)$, which should also be determined self-consistently. Note that $\Delta(q)$ in Eq.~(\ref{Eq6}) corresponds to a periodic modulated order parameter $\Delta(x)=\Delta e^{i2qx}$ in space. A nonzero Cooper pair momentum in equilibrium state indicates a generation of a helical phase (Fulde-Fellel state) \cite{Fulde,Larkin,Yuan2,LukasSciPost2020}. Note that there is also a constant in Eq.~(\ref{Eq6}) arising from the mean-field approximation for BCS mean-field Hamiltonian: $\mathrm{const}=\sum_{k}E_{-k+q,-}+\frac{\mathcal{V}}{U_{s}}\Delta^{2}(q)$. It is neglected in Eq.~(\ref{Eq6}) since it does not affect the following self-consistent calculation. Using Bogoliubov-de-Gennes
(BdG) transformation, the Hamiltonian in Eq.~(\ref{Eq6}) for every $q$ be diagonalized as: $H(q)=\sum_{k}\tilde{E}_{+}(k,q)\alpha_{k+q}^{\dagger}\alpha_{k+q}+\tilde{E}_{-}(k,q)\beta_{-k+q}\beta_{-k+q}^{\dagger}$ with the eigenvalues $\tilde{E}_{\pm}(k,q)=E_{1}(k,q)\pm \sqrt{E^{2}_{2}(k,q)+\Delta^{2}(q)}$ and $E_{1,2}(k,q)=\frac{E_{k+q,+}\mp E_{-k+q,-}}{2}$. For every fixed $q$, $\Delta(q)$ should be self-consistently determined by a gap equation \cite{Daido}:
\begin{equation}
	\begin{aligned}
	&\Delta(q)=\frac{U_{s}}{\mathcal{V}}\sum_{k}\langle c_{-k+q,-}c_{k+q,+}\rangle\\
	&=-\frac{U_{s}}{\mathcal{V}}\sum_{k}\frac{\Delta(q)}{2\sqrt{E^{2}_{2}(k,q)+\Delta^{2}(q)}}(\langle \alpha_{k+q}^{\dagger}\alpha_{k+q}\rangle-\langle \beta_{-k+q}\beta^{\dagger}_{-k+q}\rangle) \\
	&=-\frac{U_{s}}{\mathcal{V}}\sum_{k}\frac{\Delta(q)}{2\sqrt{E^{2}_{2}(k,q)+\Delta^{2}(q)}}[f(\tilde{E}_{+}(k,q))-f(\tilde{E}_{-}(k,q))].
	\end{aligned}
	\label{Eq7}
\end{equation}
Based on $\Delta(q)$ in Eq.~(\ref{Eq7}), we can calculate the free energy $\Omega(q)$ per volume:
\begin{equation}
	\Omega(q)=\frac{\Delta^{2}(q)}{U_{s}}+\frac{1}{\mathcal{V}}\sum_{k}E_{-k+q,-}
	-\frac{T}{\mathcal{V}}\sum_{k,\eta=\pm}\ln(1+e^{\frac{-\tilde{E}_{\eta}(k,q)}{T}}).
	\label{Eq9}
\end{equation}
Following the previous derivations \cite{Daido}, the superconducting current flowing through the system $j_{s}$ is evaluated as:
\begin{equation}
	\begin{aligned}
	j_{s}(\Delta(q),q)&=\frac{e}{\hbar}\partial_{q}\Omega(\Delta(q),q)\\
	&=\frac{e}{\hbar}\partial_{q}[\Omega(\Delta(q),q)-\Omega(\Delta(q)=0,q=0)]\\
	&=\frac{e}{\hbar}\partial_{q}F_{s}(q).
    \end{aligned}
	\label{Eq10}
\end{equation}
$F_{s}(q)=\Omega(\Delta(q),q)-\Omega(\Delta(q)=0,q)$ is the condensation energy per volume to quantize the difference of free energy density between the superconducting state and the normal state.  In addition, the last equation uses the fact that $\Omega(\Delta(q)=0,q)=\Omega(\Delta(q)=0,q=0)$. Note that once $F_{s}(q)>0$, we set it as zero regarding the superconducting phase is no longer stable. Eq.~(\ref{Eq10}) actually follows the standard expression $j_{s}=-\partial_{A}\Omega$ with the gauge vector $A$ \cite{Daido, Yuan2}. In addition, the intrinsic depairing currents $\tilde{j}_{c\pm}$ just corresponds the global maximum $\tilde{j}_{c+}=\max_{q}[j_{s}(q)]$ and the global minimum $\tilde{j}_{c-}=\min_{q}[j_{s}(q)]$, respectively.

For the usual case, the depairing currents which are demanded to dissolve Cooper pairings are just equal to superconducting critical currents. No superconducting state can sustain once the applied normal current $j_{ext} > \tilde{j}_{c+}$ or $j_{ext} < \tilde{j}_{c-}$ for a definite valley polarization $h_{v}$. However, the situation becomes more complex after including the effect of the current-induced valley polarization modulation. See magenta solid lines and cyan dashed lines Fig.~\ref{FIG1}(b), as the applied current $j_{ext}$ varies, the $h_{v}$ will also change following the relation of Eq.~(\ref{Eq5}). Note that $h_{v}$ in turn affects the superconducting order parameter $\Delta (q)$ as well as corresponding depairing currents $\tilde{j}_{c\pm}(h_{v})$ (red and dark blue solid lines). Therefore, the relations between $j_{ext}$ and $\tilde{j}_{c\pm}$ should now be reevaluated. In Fig.~\ref{FIG1}(b), we use dark green stars to denote intersection points between the $h_{v}-j_{ext}$ line and the $\tilde{j}_{c\pm}-h_{v}$ lines. At the regions between these intersection points (cyan dashed lines), $|j_{ext}|$ is found to be always smaller than $|\tilde{j}_{c\pm}|$, indicating that the system should stay in the superconducting phase. While in the other regions (magenta solid lines), $j_{ext} > \tilde{j}_{c+}$ or $j_{ext} < \tilde{j}_{c-}$, meaning that the normal phase should be favored. Therefore, the intersection points can define actual critical currents $j_{c \pm}$ in metal-superconductor transitions.  Moreover, we can find the characteristics of $j_{c\pm}$ strongly depends on the initial valley splitting field $h_{v}^{0}$ (intersections with the vertical axis). One notable example that the system stays in the normal phase at $A$ point (magenta region) with $j_{ext}=0$, but is driven into a superconductor (cyan region) after acrossing $B$ point with $j_{ext} > 0$. This will lead to two positive critical currents ($j_{c\pm} > 0$), which is quite similar to extreme nonreciprocity observed in previous experiment \cite{Lin}.

One point may be noticed that the physical picture shown in Fig.~\ref{FIG1}(a) will not hold when the system has been driven into the superconducting phase (cyan regions) where the electron distribution is equilibrium. Actually, our work focuses on the process of driving the system into superconductivity with a normal current. The intersection points in Fig.~\ref{FIG1}(b) are still reasonable to define critical currents for the phase transition process starting from normal phases to the superconducting phases [denoted by the arrows in Fig.~\ref{FIG1}(b)]. Once crossing the intersection points (e.g., $B$ point) from the magenta regions to cyan regions, the system cannot remain in the normal phase; otherwise, the normal current $j_{ext}$ has to continue to weaken $h_{v}$ along cyan dashed lines. At this time, the corresponding depairing currents $\tilde{j}_{c+}(h_v)$ allowed by the superconducting phase will inevitably exceed the applied normal current $J_{ext}$, indicating that the normal phase is no longer favored. This judgement does not yet involve the specific behaviors of currents and valley polarizations within the superconducting phase. 

\begin{figure*}[ht]
	\includegraphics[width=0.98\textwidth]{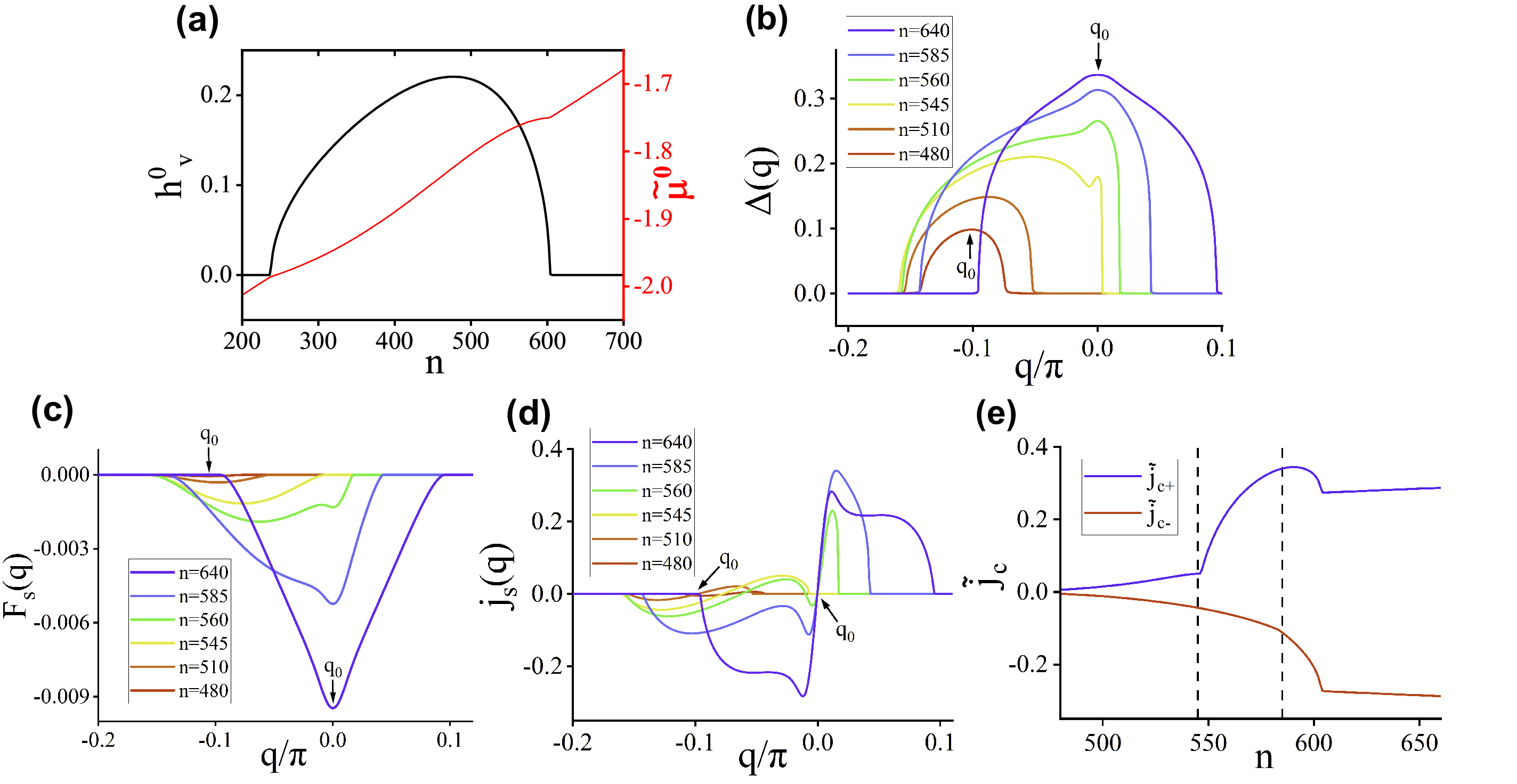}
	\centering %
	\caption{(a) The initial valley splitting field $h_{v}^{0}$
(dark line) and initial modified chemical potential
$\tilde{\mu}^0$ (red line) versus the electron occupation $n$.
(b, c, d) The distribution of superconducting order parameter $\Delta(q)$, the condensation energy $F_{s}(q)$ and supercurrent $j_{s}(q)$ for several $n$. (e) The variation of intrinsic deparing currents $\tilde{j}_{c\pm}$ versus the filling $n$.}
	\label{FIG2}
\end{figure*}

On the other hand, although the system has reached equilibrium when entering the superconducting phase, we emphasis equilibrium supercurrents can still couple to valley polarizations \cite{Banerjee}. Similar to the Fig.~\ref{FIG1}(a), the finite momentum $2q$ of Cooper pairs carrying the supercurrent will also lift or lower the energy bands according to band dispersions. For valley bands with the broken intravalley inversion symmetry, this band shifts induced by Cooper-pair momentum cannot be simply offset, still leading to the change of valley polarizations (see some discussions in Appendix.~\ref{SEC-D}). Within the superconducting phases, the interplay between supercurrents-induced valley polarization modulations and superconductivity will also modulate critical currents.

In Eq.~(\ref{Eq5}), a larger coefficient $\alpha_{+}-\alpha_{-}$ implies $j_{ext}$ can weaken $h_v$ more quickly and drive the system into superconducting phase more easily. Below we choose a simple one-dimensional (1D) two-band toy model with
$\epsilon_{k,+}=-2t\cos[\frac{8}{15}(k-\frac{7\pi}{8})]$ for $-\pi \leqslant k \leqslant \frac{7\pi}{8}$,
$\epsilon_{k,+}=-2t\cos(8k-\pi)$ for $\frac{7\pi}{8} < k < \pi$,
and $\epsilon_{k,+}=\epsilon_{-k,-}$.
Similar model is used to illustrate the interplay between spontaneous valley polarization and applied currents, and could  capture the asymmetric features of low-energy bands in twisted graphene \cite{Su}. In numerical calculations, $\mathcal{V}=Na$ with a periodic boundary condition and $N=2000$.
$t=1$, $\frac{e}{h}t=1$, and $a=1$ are set as energy,
current and length units, respectively.
We also set $U_{v}=2.8$, $U_{s}=1.86$ and thermal energy $T=0.1$. In Fig.~\ref{FIG1}(c), the coefficient $(\alpha_{+}-\alpha_{-})/N$ versus the initial $h^{0}_{v}$ and $\tilde{\mu}^0$ is shown.
$(\alpha_{+}-\alpha_{-})/N$ dives as $\tilde{\mu}^0$
becomes lower, considering Fermi velocities approach zero
and $\alpha_{\tau}$ becomes divergent near the bottom of bands. Furthermore, we also demonstrate a relatively significant $\alpha_{+}-\alpha_{-}$ coefficient exist in a more realistic tight-binding model for twisted bilayer graphene. And the more asymmetrical the bands are, the larger $\alpha_{+}-\alpha_{-}$ is (see both discussions in Appendix.~\ref{SEC-E}).

\section{\label{SEC3} Numerical results}
In this section, we will use a series of numerical calculations to validate our physical pictures illustrated in Fig.~\ref{FIG1}. We first study the initial valley splitting field $h^{0}_{v}$ and corresponding depairing currents $\tilde{j}_{c\pm}$ without the effect of current-induced valley polarization modulation, see Sec.~\ref{S31}. Then, through the remodulation process shown in Fig.~\ref{FIG1}(b), we demonstrate actual critical currents $j_{c}$ and explore the situation where the extreme nonreciprocity appears, see Sec.~\ref{S32}. We will also investigate the influence on $j_{c}$ by the electron occupation $n$ and the external magnetic field $B$, see Sec.~\ref{S33}.

\subsection{\label{S31} The calculations without the effect of current-induced valley polarizations}

For a given electron occupation $n$ (or filling factor $\nu=n/N$), the initial $\tilde{\mu}^0$ and $h_{v}^{0}$ can be solved self-consistently
from $H^{v}_{MF}$ and are shown in Fig.~\ref{FIG2}(a).
Notice that $\pm h_{v}^{0}$ are degenerate solutions
but we choose the positive one like the
magnetic training in the experiment \cite{Lin}.
Here $\tilde{\mu}^0$ naturally declines as $n$ decreases. Especially, a non-zero $h^{0}_{v}$ appears around $240<n<600$. Based on $h_{v}^0$,
the $\Delta(q)$ is solved from the self-consistent gap equation in Eq.~(\ref{Eq7}), and is shown in  Fig.~\ref{FIG2}(b). As $n$ declines from $n=640$ to $n=480$, $h^{0}_{v}$ becomes stronger and $\Delta(q)$ becomes weaker and more asymmetric with $\Delta(q)\neq \Delta(-q)$. This is because $h_{v}^{0}$ breaks TRS
and destroys the Cooper pairs from intervalley pairings.
Additionally, when $n < 545$, the strong $h_{v}^{0}$ causes
that the center of $\Delta(q)$ wholely shifts
from $q_{0}=0$ to $q_{0} \approx -0.1\pi$, which apparently suggests Cooper pairs have large non-zero center of mass momenta [Fig.~\ref{FIG2}(b)].

Based on $\Delta(q)$, we also calculate corresponding condensation energy density $F_{s}(q)$, as shown in Fig.~\ref{FIG2}(c). As $n$ changes from $n=640$ to $n=480$, the initial valley splitting field increases to break TRS and intervalley pairing, thus $F_{s}(q)$ becomes much more asymmetric and narrower. At around $n=480$, $F_{s}(q)$ reaches almost zero and indicates superconductivity is highly unstable. Specially, a single-well structure of $F_{s}(q)$ with one global minimum assigning the ground state at $q_{0}=0$ ($n=640$) gradually evolves into a double-well structure under a moderate valley splitting field (e.g. $n=560$) with two local minimums. It goes back to the single-well structure with one minimum at $q_{0} \approx -0.1\pi$ under a high valley splitting field (e.g. $n=545$). Overall, the shift of the minimum point for $F_{s}(q)$ implies the superconductor transforms from a `weak' helical phase to a `strong' helical phase as the valley splitting field climbs \cite{Ilic}. 

\begin{figure}[ht]	\includegraphics[width=0.85\columnwidth]{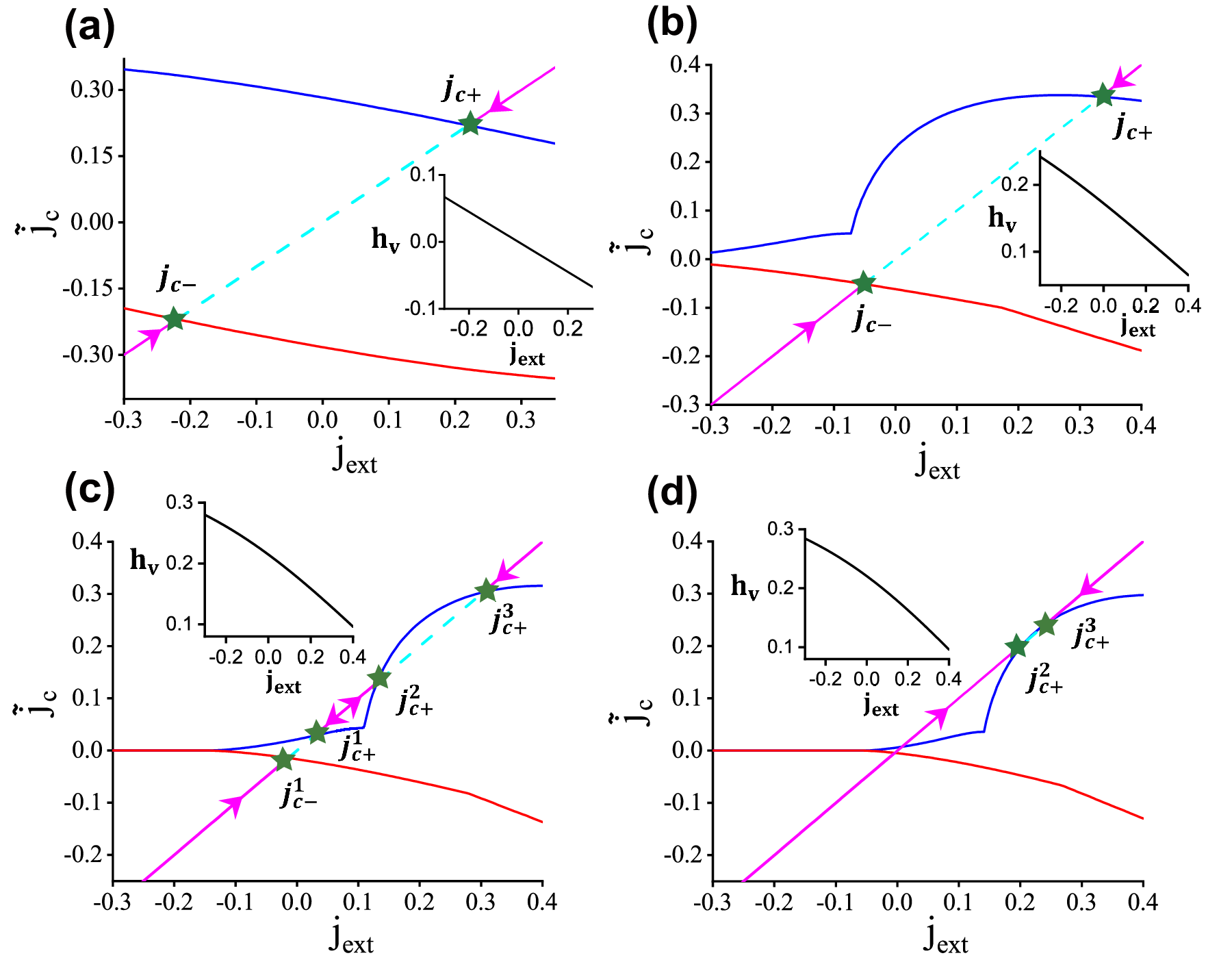}
	\centering
	\caption{(a-d) The intrinsic depairing currents $\tilde{j}_c$ (main panels) and $h_v$ versus $j_{ext}$ (insets) for $n = 640$ (a), $n = 560$ (b), $n = 510$ (c) and $n = 480$ (d). The intersection points (dark green stars) between the $\tilde{j}_{c,\pm}-j_{ext}$ (red and dark blue solid lines) and $\tilde{j}_c = j_{ext}$ (magenta solid lines and cyan dashed lines) are $j_{c,\pm}$. Similar to Fig.~\ref{FIG1}(b), the magenta parts denote the regions of the normal phases and the cyan parts denote the regions where the systems eventually transition into superconducting phases. The magenta arrows denote the phase transition from normal states to superconducting states that our theory focuses.}
	\label{FIG3}
\end{figure}

We further estimate supercurrents $j_{s}(q)$ and depairing currents $\tilde{j}_{c\pm}$ versus $n$ [Figs.~\ref{FIG2}(d, e)].  For about $n>600$, $j_{s}$ appears as an odd function with $\tilde{j}_{c+}=-\tilde{j}_{c-}$ since the initial valley splitting field $h_{v}^{0}$ is zero [Fig.~\ref{FIG2}(a)]. As $n$ decreases, $h_{v}^{0}$ climbs and $j_{s}(q)$ becomes asymmetrical. $|\tilde{j}_{c-}|$ gradually decays while $\tilde{j}_{c+}$ lifts slightly because a small $h_{v}^{0}$ gives Cooper pairs finite momenta to flow towards one direction more easily [Fig.~\ref{FIG2}(e)]. By further decreasing $n$, two additional local extrema appear in $j_{s}(q)$ around a relatively high momentum $q_{0} \approx -0.1\pi$ [Fig.~\ref{FIG2}(d)], and they successively become the new global minimum $\tilde{j}_{c-}$ ($n<585$) and maximum $\tilde{j}_{c+}$ ($n<545$) [denoted by black dashed lines in Fig.~\ref{FIG2}(e)]. Especially, the difference between $\tilde{j}_{c\pm}$ appears to be tiny after a transition from the `weak' helical phase in low $h_{v}^{0}$ to the `strong' helical phase in high $h_{v}^{0}$, see Figs.~\ref{FIG2}(c,d).

\subsection{\label{S32} The actual critical currents through the remodulation process}

Including the effect of current-induced valley polarization modulation, we state that depairing currents $\tilde{j}_c$ can be further remodulated as the actual critical currents $j_{c}$. As illustrated in Fig.~\ref{FIG1}(b), $j_{c}$ can be determined by intersection points between the curve $\tilde{j}_{c}(h_{v})$ and the curve $h_{v}(j_{ext})$. Note that $j_{ext}$ and $h_{v}$ have a definite relation in Eq.~(\ref{Eq5}). Equivalently, we show diagrams with curves $\tilde{j}_{c, \pm}(j_{ext})$ (colored solid lines) and curves $\tilde{j}_{c}=j_{ext}$ (colored dashed lines) for four different $n$ in Fig.~\ref{FIG3}.

\begin{figure}[ht]
	\includegraphics[width=0.9\columnwidth]{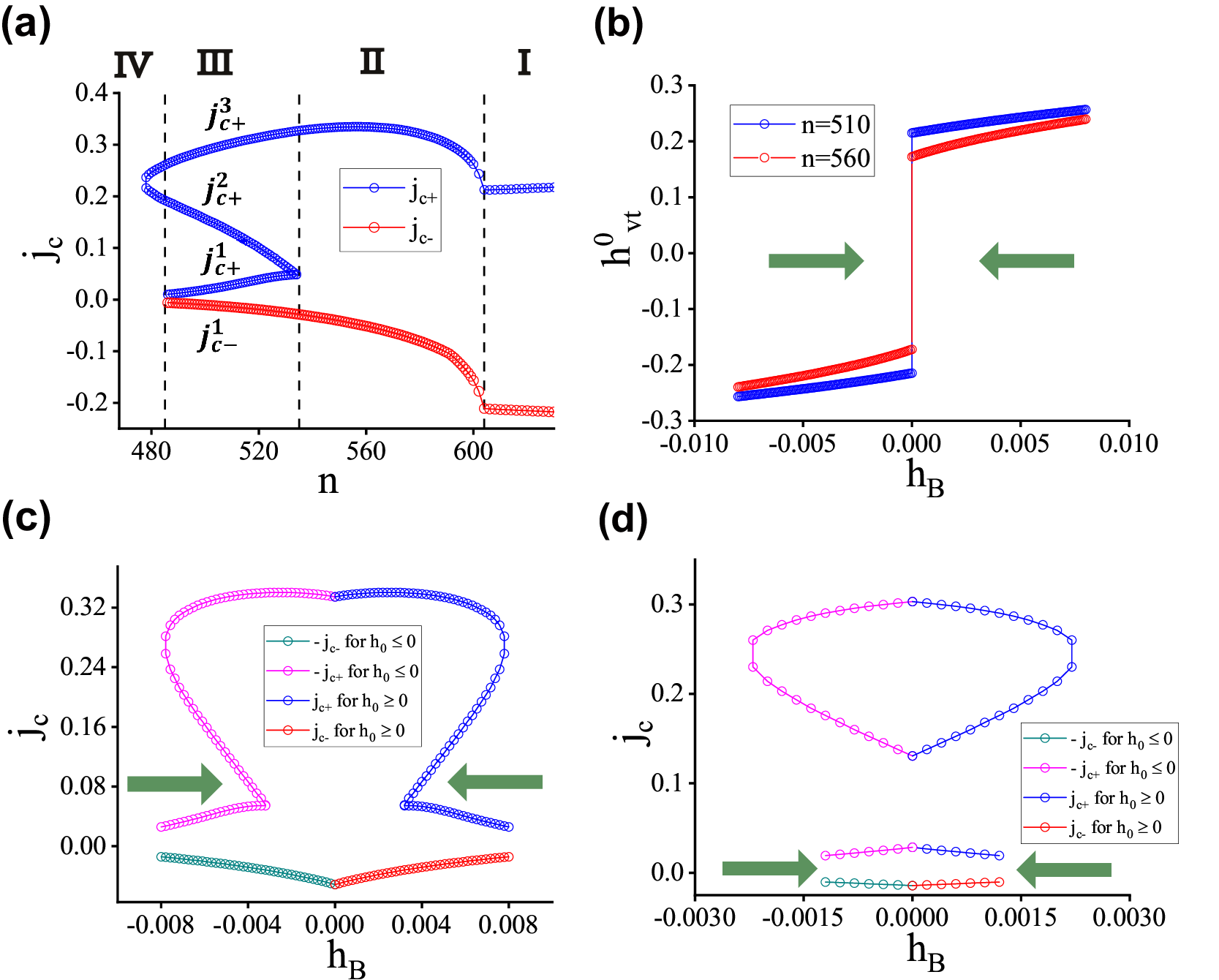}
			\centering
			\caption{(a) The variation of $j_{c \pm}$ as a function of $n$. (b) The modulation of the total valley splitting field $h_{vt}^0$ with $h_{B}$ induced by the external magnetic field. (c,d) $j_{c \pm}$ versus $h_{B}$ for $n=560$ (c) and $n=510$ (d) \cite{shuoming}. The dark green arrows denote the scanning directions of the magnetic field $B$ or $h_{B}$.}
		\label{FIG4}
		\end{figure}

In Fig.~\ref{FIG3}(a) with $n=640$, $h^0_v$ is zero and
depairing currents satisfy $\tilde{j}_{c+}=-\tilde{j}_{c-}$ at $j_{ext}=0$. A non-zero applied current $j_{ext}$ can evolve $h_{v}$ to finite
[inset in Fig.~\ref{FIG3}(a)], and simultaneously affect $\tilde{j}_{c\pm}$. While, intersection points still satisfy $j_{c+}=-j_{c-}$ indicating no SDE (dark green stars), due to the fact that $\tilde{j}_{c+}(h_{v})=-\tilde{j}_{c-}(-h_{v})$ and $h_{v}(j_{ext})=-h_{v}(-j_{ext})$ at this case. When $n=560$, a small $h_{v}^{0}$ appears and $\tilde{j}_{c+} \neq |\tilde{j}_{c-}|$ at $j_{ext}=0$ in Fig.~\ref{FIG3}(b). The forward current ($j_{ext}>0$) reduces $h_{v}$ while the backward current ($j_{ext}<0$) enhances the $h_{v}$[inset in Fig.~\ref{FIG3}(b)]. SDE persists with two modified $j_{c\pm}$ (dark green stars). When $n=510$ [Fig.~\ref{FIG3}(c)], $h_{v}^{0}$ is relatively strong and the system enters a `strong' helical superconducting phase as indicated by Figs.~\ref{FIG2}(c,d). The depairing currents $\tilde{j}_{c\pm}$ are relatively small at $j_{ext}=0$.

Interestingly, since there is a sudden change in the slope of curve $\tilde{j}_{c+}(j_{ext}$) [as indicated in Fig.~\ref{FIG2}(e)], the number of intersection points could be four, which are symbolized by four actual critical currents ($j^1_{c-}$ and $j^{1-3}_{c+}$). Regarding that there exist two different superconducting phases with two distinct critical current intervals \cite{Cao}, We call this phenomenon as current-induced re-entrant superconductivity.  Once $h_{v}^{0}$ becomes too large [see Fig.~\ref{FIG3}(d) with $n=480$], $j^1_{c\pm}$ obviously shrink towards zero and hard to be measured in the experiment, while $j^{2,3}_{c+}$ persist. Now it exhibits the extreme nonreciprocity only with two positive actual critical currents.

\begin{figure*}[ht]
	\includegraphics[width=1.02\textwidth]{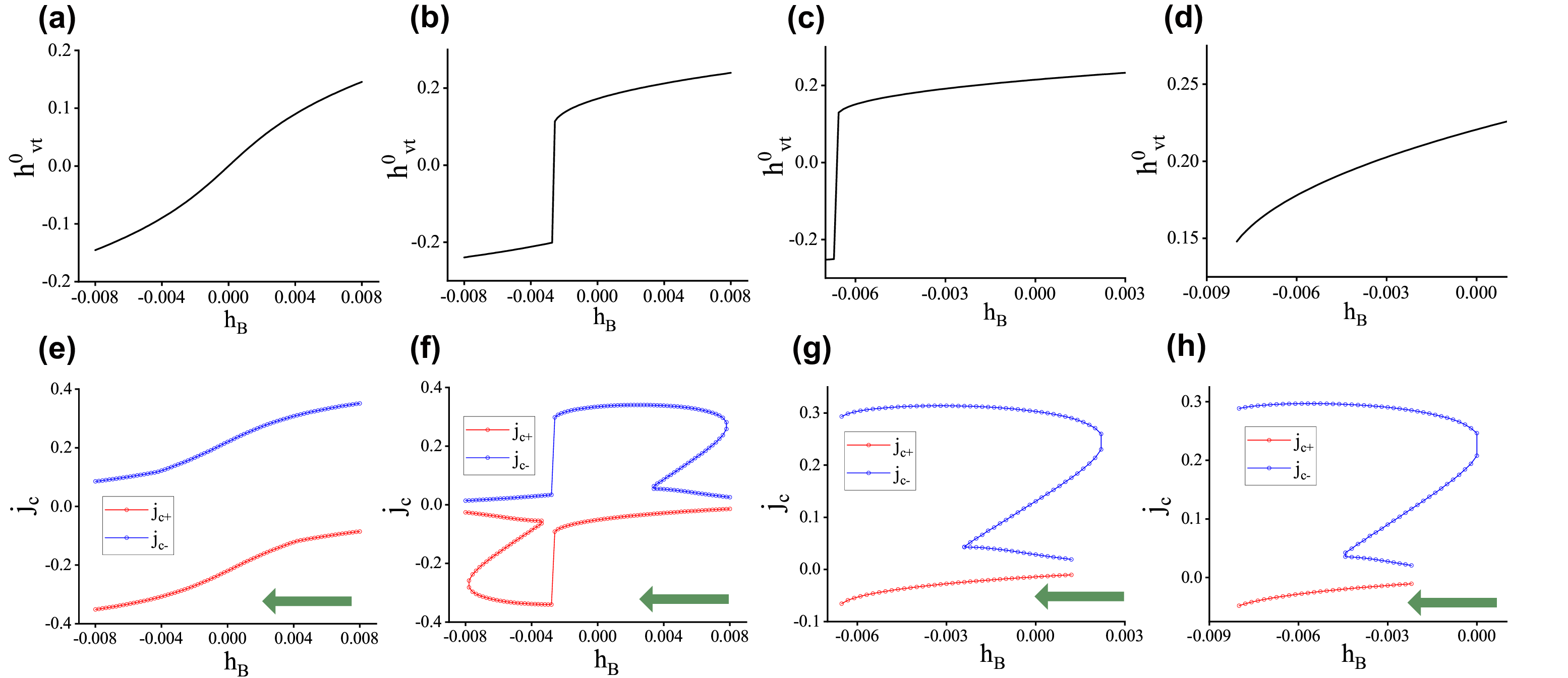}
	\centering %
	\caption{(a-d) The change of the total valley splitting field $h_{vt}^{0}$ as a function of the $h_{B}$ for electron occupation $n=640$ (a), $n=560$ (b), $n=510$ (c) and $n=480$ (d). The system is initially prepared at the stable state at $m \geq 0$ without the external magnetic field $B$. And the magnetic field as well as the additional valley splitting field $h_{B}$ is scanned from positive to negative which influences the self-consistent result $h_{vt}^{0}$ in every step. (e-h) the change of the actual critical currents $j_{c}$ as a function of $h_{B}$, corresponding to the cases in (a-d), respectively. The dark green arrows denote the scanning direction.}
	\label{FIG5}
\end{figure*}

Similar to the schematic diagram in Fig.~\ref{FIG1}(b), we use magenta solid lines to mark normal phase regions and use cyan dashed lines to denote superconducting regions in Fig.~\ref{FIG3}. Following the magenta arrows, when the system is initially prepared in normal phases (magenta regions) and driven by the normal current to cross intersection points, the normal phase cannot be maintained, otherwise $h_v$ will continue to be weakened along the cyan dashed line. Our theory focuses on the current-induced phase transition from normal phases to superconducting phases and offers a possible mechanism for the observation of extreme nonreciprocal SDE in ref.\cite{Lin}. Additionally, when the system stays in the superconducting phase, the supercurrents will still couple to supercurrents \cite{Banerjee}. As the supercurrent-induced valley polarization modulation is not completely the same as the normal current case, there may be a hysteresis behaviour when the system in turn transitions from the superconducting phase and the normal phase.

\subsection{\label{S33} The variation of actual critical currents with different parameters}

To study the SDE comprehensively, in Fig.~\ref{FIG4}(a)
we extract $j_{c\pm}$ based on the intersection points in Fig.~\ref{FIG3} and show them in a wide range of electron occupations $n$. Here four regions are denoted.
In region \textbf{I}, the system does not exhibit SDE due to zero $h_{v}^{0}$. In region \textbf{II}, $h_{v}^{0}$ is moderate and the conventional SDE with $j_{c+} \neq |j_{c-}|$ is observed. $j_{c+} - |j_{c-}|$ becomes roughly larger as $n$ decreases ($h_{v}^{0}$ climbs). In region \textbf{III}, $h_{v}^{0}$ is relatively large. The system exhibits re-entrant superconductivity with four actual critical currents ($j^1_{c-}$ and $j^{1-3}_{c+}$). In region \textbf{IV}, $h_{v}^{0}$ is stronger and $j^{1}_{c\pm}$ become too small to be observed.
Only $j^{2,3}_{c+}$  are left and thus the system exhibits an obvious extreme nonreciprocity. When $h_{v}^{0}$ becomes too large ($n$ is small), both $j^{2,3}_{c+}$ will disappear and the superconducting phase cannot exist. The extreme nonreciprocity occurs near the disappearance of superconductivity in our theory is akin the feature in ref \cite{Lin}. 

Besides varying fillings, we also investigate how an external magnetic field $B$ can modulate $j_{c\pm}$. Enlightened by ref.~\cite{Lin}, we now consider the valley $\tau$ is locked with the spin $s_{z}$, which can arise from the Ising spin-orbit coupling \cite{Xiao, Zhou2}. Thus, $B$ can couple to valley through a Zeeman effect and induce an additional valley splitting field $h_{B}\varpropto B$
into the Hamiltonian $H^v_{MF}=\sum_{k,\tau}(E_{k,\tau}-h_{B}\tau)c^{\dagger}_{k,\tau}c_{k,\tau}$. Through similar self-consistent calculations in Eq.~(\ref{Eq3}), the total valley splitting field $h^{0}_{vt}=h^{0}_{v}+ h_{B}$ is refreshed along with the magnetic field.

In Fig.~\ref{FIG4}(b), we plot the calculated $h^{0}_{vt}$ versus $h_{B}$. Note that here the system is initially prepared at $h^{0}_{v}>0$ ($h^{0}_{v}<0$) before applying the magnetic field $B>0$ ($B<0$). It roughly corresponds to magnetic field $B$ scanned from positive (negative) direction to zero (see dark green arrows). Actually, these two cases are antisymmetric due to TRS. We can find $|h_{vt}^{0}|$ decays as $|h_{B}|$ weakens, reflecting the modulation of valley polarizations by the external magnetic field. We also plot $j_{c}$ versus $h_{B}$ for two distinct $n$. For $n=560$ in Fig.~\ref{FIG4}(c), as $h_{B}$ sweeps from positive to zero, the decay of $h_{vt}^{0}$ drives the number of actual critical currents from 4 to 2. It means, the system evolves from a re-entrant superconducting phase to a conventional SDE. For $n=510$ in Fig.~\ref{FIG4}(d), the system is initially an extreme nonreciprocal SDE with two positive $j_{c}$ at $|h_{B}| \approx 0.002$. The decline of $|h_{B}|$ pulls down $h_{vt}^{0}$ and impels the system into re-entrant superconducting phase with four distinct $j_{c}$. 

Additionally, the polarity of SDE may be also reversed when scanning $B$ from the positive to negative direction, see Fig.~\ref{FIG5}. At these cases, the system is initially prepared at $m \geq 0$ for $n=640$ [Figs.~\ref{FIG5}(a,e)], $n=560$ [Figs.~\ref{FIG5}(b,f)], $n=510$ [Figs.~\ref{FIG5}(c,g)] and $n=480$ [Figs.~\ref{FIG5}(d,h)], respectively. Then, we apply and scan $h_{B} \propto  B$ from positive to negative (dark green arrows). For $n=640$, since no valley polarization appears without the external magnetic field ($h_{B}=0$), both the sign of $h^{0}_{vt}$ and the polarity of SDE correlates well with $h_{B}$ [Figs.~\ref{FIG5}(a,e)]. For the case of $n=560$ and $n=510$, a sudden sign change of $h_{vt}^{0}$ appears as $h_{B}$ reaches about $-0.0027$ [Fig.~\ref{FIG5}(b)] and $-0.007$ [Fig.~\ref{FIG5}(c)]. This switching could also reverse the polarity of SDE in Fig.~\ref{FIG5}(f). Note that the switching of $h_{vt}^{0}$ in Fig.~\ref{FIG5}(c) is so large that $j_{c\pm}$ disappears in Fig.~\ref{FIG5}(g). Similar to Fig.~\ref{FIG4}, the number of $j_{c}$ also varies with $h_{B}$ which manifests the transformation of types of SDE [Figs.~\ref{FIG5}(f,g)]. Since the initial valley splitting field $h_{v}^{0}$ for $h_{B} = 0$ is too large for $n=480$, a small variation of magnetic field is not enough to switch total valley splitting field $h^{0}_{vt}$ [Fig.~\ref{FIG5}(d)]. But the superconducting phase gradually transforms from an extreme nonreciprocity with two positive $j_{c}$ to the re-entrant superconductivity with four $j_{c}$ and then to the conventional SDE with $j_{c+}>0 , j_{c-}<0$ [Fig.~\ref{FIG5}(h)]. 

In summary, our results in Fig.~\ref{FIG4} and Fig.~\ref{FIG5} both demonstrate that the extreme nonreciprocity can occur and be adjusted by the variation of the electron occupation $n$ and the external magnetic field $B$. Additionally, our results are robust to changes in system size (see Appendix.~\ref{SEC-F}).

\section{\label{SEC4} Discussions and conclusion}

The toy model we calculated can qualitatively explain the phenomena observed in experiments. In fact, through a rough estimation, we also find that the calculated results are also similar in magnitude to experimental measurements. Considering a narrow bandwidth of the flat band with $4t=10$ meV \cite{Cao2,Cao3,Codecido,Phong}, the energy unit becomes $t=2.5$ meV and the current unit becomes $\frac{e}{h}t \approx 96.6$ nA. Thus, the set temperature $T$ in the unit of Kelvin is around 2.9 K. In Fig.~\ref{FIG2}, the initial valley splitting field $h_{v}^{0}$ varies from 0 to about $0.2t$ (0.5 meV) and the maximal superconducting order parameter is $\Delta \approx 0.33t \approx 0.83$ meV. In Fig.~\ref{FIG4}(a), we can roughly estimate the amplitudes of critical currents $j_{c\pm}$ varying from 0 to 34 nA, which is in order of magnitude consistent with the previous experiment results \cite{Lin}. It is also worth noting that the normal currents arranging from several nA to tens of nA is experimentally confirmed to be able to affect magnetizations in twisted bilayer graphene \cite{Serlin, Sharpe}, which are still similar in magnitudes for current-induced valley polarization modulations in our theoretical scheme. Totally speaking, the modulation of valley splitting field caused by the weak current is not strong in our results. See Fig.~\ref{FIG3}, $h_{v}$ changes by about $0.1t$ (about 0.25 meV) as the current changes by about 0.4$\frac{e}{h}t$ (about 40 nA). 

In conclusion, based on a simple valley-polarized model, we have revealed that intrinsic depairing currents can be remodulated due to the current-induced valley polarization modulation. Depending on specific features, we have demonstrated that such a remodulation can induce the extreme nonreciprocity and also the current-induced re-entrant superconductivity. These special SDE can be further adjusted by varing electron occupations and external magnetic fields. Our study reflects the peculiarity in the interplay between valley ferromagnetism and superconductivity, provides a possible mechanism to explain experimental observations of extreme nonreciprocal SDE and opens a new way to implement SDE with 100\% efficiency. 

\section*{Acknowledgements}
We are grateful to Yue Mao and Yi-Xin Dai for fruitful discussions.

% TODO: include author contributions
%\paragraph{Author contributions}
%This is optional. If desired, contributions should be succinctly described in a single short paragraph, using author initials.

% TODO: include funding information
\paragraph{Funding information}
This work was financially supported by 
the National Natural Science Foundation of China (Grants No. 12447146, No. 12374034 and No. 11921005), the National Key R and D Program of China (Grant No. 2024YFA1409002), Quantum Science and Technology-National Science and Technology Major project (2021ZD0302403), the China Postdoctoral Science Foundation (No. 2025T180938) and the Postdoctoral Fellowship Program of CPSF under Grant No. GZB20240031. We also acknowledge the High-performance Computing Platform of Peking University for providing computational resources.\\

\begin{figure}[ht]
	\includegraphics[width=0.85\columnwidth]{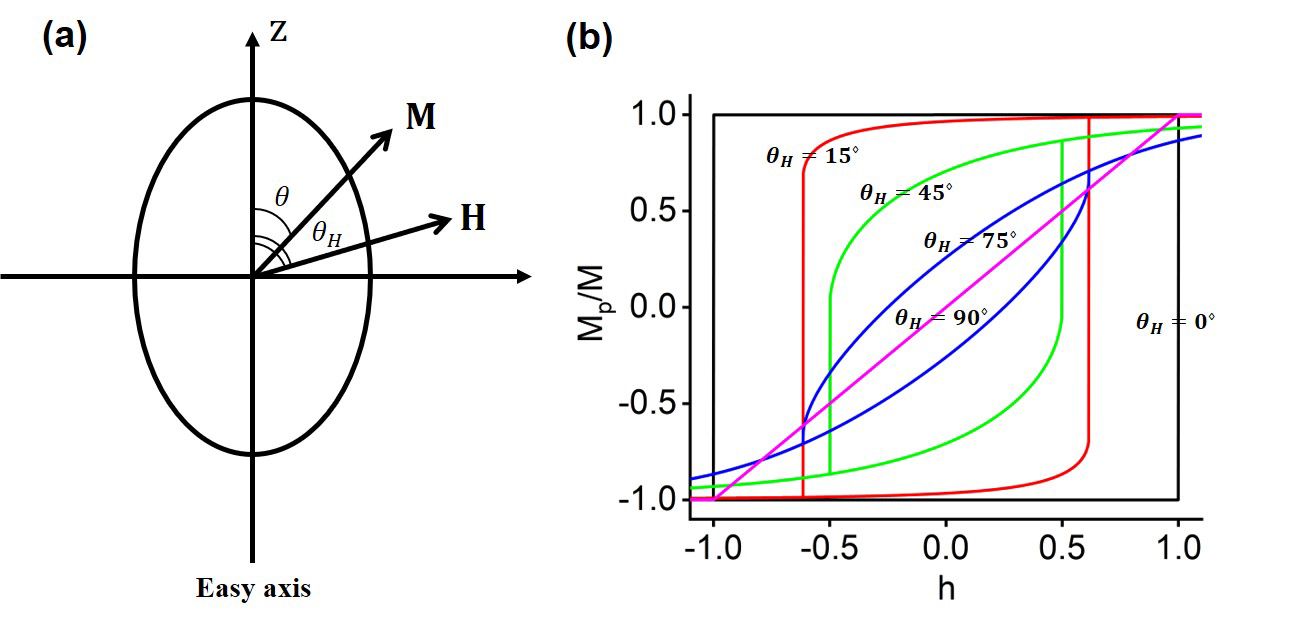}
		\centering %
		\caption{(a) The model for a single-domain spheroidal particle endowed with uniaxial magnetic anisotropy. The magnetization M is aligned with an angle $\theta$ between the easy axis z and the external magnetic field H is aligned with an angle $\theta_H$ between the easy axis z. (b) The numerical calculated magnetization curves between $\mathrm{M_p/M}$ and h for different angles $\theta_H$.}
		\label{FIG6}
\end{figure}

\begin{appendix}
\numberwithin{equation}{section}
\section{\label{SEC-A}Formulations of the current-induced valley polarization modulation}

When the applied current $j_{ext}$ flows through the valley-polarized system shown in Eq.~(\ref{Eq1}), occupations for electrons with opposite group velocities should be further imbalanced. In detail, taking a 1D system shown in Fig.~\ref{FIG1}(a) as an example with $\mathcal{V}=Na$ where $a=1$ is the length unit, the Fermi level of electrons with positive (negative) velocities, coming from the source (drain) will rise (fall) $\frac{eV}{2}$, respectively. Then, the electron occupation $n_{\tau}$ on each valley $\tau$ changes into:
\begin{equation}	
	n_{\tau}=\sum_{k}f[\epsilon_{k,\tau}-\tilde{\mu}-h_{v}\tau-\frac{eV}{2}\mathrm{sgn}(\epsilon'_{k,\tau})]
	\label{EqA1}
\end{equation}
with $\mathrm{sgn}(x>0)=1, \mathrm{sgn}(x=0)=0, \mathrm{sgn}(x<0)=-1$. And $e$ is the electron charge.
For a small bias $V \rightarrow 0$, Eq.~(\ref{EqA1}) can be further approximated as:
\begin{equation}
n_{\tau}\approx n_{\tau}^{0}-\frac{eV}{2}\sum_{k}\mathrm{sgn}(\epsilon'_{k,\tau})f'(\epsilon_{k,\tau}-\tilde{\mu}-h_{v}\tau),
\label{EqA2}
\end{equation}
where $n^{0}_{\tau}=\sum_{k}f(\epsilon_{k,\tau}-\tilde{\mu}-h_{v}\tau)$ is the original electron occupation before applying the current $j_{ext}$. Furthermore, the current $j_{ext}$ flowing through the system which is closely related to the voltage $V$ can be also calculated:
\begin{equation}
	\begin{aligned}
	j_{ext}&=\frac{e}{\hbar N}\sum_{k,\tau}\epsilon'_{k,\tau}f[\epsilon_{k,\tau}-\tilde{\mu}-h_{v}\tau-\frac{eV}{2}\mathrm{sgn}(\epsilon'_{k,\tau})] \\
	&\approx \frac{e}{h}\int dk \sum_{\tau}\epsilon'_{k,\tau} f[\epsilon_{k,\tau}-\tilde{\mu}-h_{v}\tau-\frac{eV}{2}\mathrm{sgn}(\epsilon'_{k,\tau})].
    \end{aligned}
	\label{EqA3}
\end{equation}
Here we regard the $N \to \infty$ and thus the summation of $k$ changes into the integral for simplicity. Similarly, when the bias is small with $V \rightarrow 0$, the current in Eq.~(\ref{EqA3}) can be approximated as:
\begin{equation}
	j_{ext} \approx \frac{e^2V}{h}\sum_{\tau}[-f(\epsilon_{k,\tau}^{max}-\tilde{\mu}-h_{v}\tau)+f(\epsilon_{k,\tau}^{min}-\tilde{\mu}-h_{v}\tau)],
\label{EqA4}
\end{equation}
where $\epsilon^{max}_{k,\tau}$ and $\epsilon^{min}_{k,\tau}$ is the global maximum and minimum value of $\epsilon_{k,\tau}$. For a low temperature $T \rightarrow 0$ and $\tilde{\mu} \in (\epsilon^{min}_{k,\tau}-h_{v}\tau,\epsilon^{max}_{k,\tau}-h_{v}\tau)$, $j_{ext}=\frac{2e^{2}V}{h}$ well corresponds to Landauer-B$\ddot{u}$ttiker formula in a ballistic regime \cite{Datta}. Substituting Eq.~(\ref{EqA4}) into Eq.~(\ref{EqA2}), we can get the relation between $n_{\tau}$ and $j_{ext}$ as
\begin{equation}
	n_{\tau}=n^{0}_{\tau}+\alpha_{\tau}j_{ext},
	\label{EqA5}
\end{equation}
and also the change of valley splitting field $h_{v}$:
\begin{equation}
h_{v}=\frac{U_{v}}{2\mathcal{V}}(n_{+}-n_{-})=\frac{U_{v}}{2\mathcal{V}}(\alpha_{+}-\alpha_{-})j_{ext}+h_{v}^{0}
	\label{EqA6}
\end{equation}
where $h_{v}^{0}$ is the initial valley splitting field when $j_{ext}=0$. The Eq.~(\ref{EqA6}) is just Eq.~(\ref{Eq5}) in the main text. The coefficient $\alpha_{\tau}$ to measure the ability for the current to modulate the valley polarization is a function of modified chemical potential $\tilde{\mu}$ and valley splitting field $h_{v}$:
\begin{equation}
	\begin{aligned}
	&\alpha_{\tau}(\tilde{\mu},h_{v})\\
	&=\frac{h\sum_{k}\mathrm{sgn}(\epsilon'_{k,\tau})f'(\epsilon_{k,\tau}-\tilde{\mu}-h_{v}\tau)}{2e\sum_{\tau}[f(\epsilon_{k,\tau}^{max}-\tilde{\mu}-h_{v}\tau)-f(\epsilon_{k,\tau}^{min}-\tilde{\mu}-h_{v}\tau)]}.
    \end{aligned}
    \label{EqA7}
\end{equation}
During applying an electric current $j_{ext}$, the change of $h_{v}$ and $\tilde{\mu}$ could alter the value of $\alpha_{\tau}$ in time. For simplicity, we ignore this effect and directly set $\alpha_{\tau}(\tilde{\mu},h_{v})$ as  $\alpha_{\tau}(\tilde{\mu}^0,h^0_{v})$. The detailed distribution of  $\alpha_{\tau}(\tilde{\mu}^0,h^0_{v})$ for an 1D toy model in Sec.~\ref{S23} is shown in Fig.~\ref{FIG1}(c). Note that its unit is $h/et$.

Especially when $T \rightarrow 0$, $\alpha_{\tau} \propto \sum_{n} \frac{1}{\epsilon'_{\tau}(k_{F}^{n})}$ where $k^{n}_{F}$ is the $n$-th Fermi wave vector at the Fermi level $E_{f}$. Thus, the value of $\alpha_{\tau}$ is closely related to the inverse of Fermi velocities $v^{n}_{F}=\frac{1}{\hbar}\epsilon'_{\tau}(k_{F}^{n})$. If the energy band has the intravalley inversion symmetry: $\epsilon_{k,\tau}=\epsilon_{-k,\tau}$. It leads to $\epsilon'_{k,\tau}=-\epsilon'_{-k,\tau}$ and $\alpha_{\tau}$ as well as the change in Eq.~(\ref{EqA6}) should be canceled to be zero. The necessity of intravalley inversion breaking is consistent with the finding in ref.~\cite{Su}. In addition, the intravalley inversion symmetry breaking is also found as a crucial condition to realize the SDE in the valley polarized system \cite{Hu}. This coincidence implies the possibility for the combination between the current-induced valley polarization modulation and SDE.

In numerical calculations, we do not directly use the $\alpha_{\tau}$ to obtain the results of current-induced valley modulation in Eq.~(\ref{EqA6}). To be more accurate, after given the applied current $I_{ext}$, we use Eq.~(\ref{EqA4}) to obtain the corresponding bias energy $eV$. Then we bring $eV$ into Eq.~(\ref{EqA1}) to get the electron occupation $n_{\tau}$ on each valley and also use $h_{v}=\frac{U_v}{2\mathcal{V}}(n_{+}-n_{-})$ to obtain the corresponding $h_{v}$. Note that the modified chemical potential $\tilde{\mu}$ and the valley splitting field $h_{v}$ are fixed as $\tilde{\mu}^0$ and $h^0_{v}$ in the right parts in Eq.~(\ref{EqA1})$-$Eq.~(\ref{EqA4}). 
In Eq.~(\ref{EqA1}), we do not assume that the bias $eV$ is small, so that $h_{v}=\frac{U_{v}}{2\mathcal{V}}(n_{+}-n_{-})$ versus the bias $eV$ deviates slightly from the linear relation (see the insets of Fig.~\ref{FIG3}).

\section{\label{SEC-B}The self-consistent manner including the effect of applied currents}

In the main text, we set the total valley polarization in Eq.~(\ref{Eq5}) is the summation of the current-induced part and the spontaneous polarization part from the Coulomb interaction. The influence on $h_{v}^{0}$ by the applied current $j_{ext}$ is just neglected. In this Appendix, we will discuss this self-consistent process theoretically and demonstrate the rationality of our linear approximation in Eq.~(\ref{Eq5}) for a small $j_{ext}$.

Actually, Eq.~(\ref{Eq5}) is easy to recall from the relationship between the magnetic induction $\mathbf{B}$ and magnetic ﬁeld strength $\mathbf{H}$:
\begin{equation}
\mathbf{B}/\mu_0 = \mathbf{M} + \mathbf{H}.
\label{EqB1}
\end{equation}
where $\mathbf{M}$ is the magnetization and $\mu_0$ is the permeability of free space. Neglecting the coefficients, $j_{ext}$, $h^0_{v}$ and $h_{v}$ in Eq.~(\ref{Eq5}) just corresponds to $\mathbf{H}$, $\mathbf{M}$ and $\mathbf{B}$ in Eq.~(\ref{EqB1}), respectively. In the magnetization process, the external magnetic ﬁeld strength $\mathbf{H}$ could also affect the intrinsic magnetization $\mathbf{M}(\mathbf{H})$, causing the relationship between $\mathbf{B}$ and $\mathbf{H}$ more complicated, usually along with magnetic hysteresis loops. Based on a rough analogy, we can draw on the magnetic curve of $M = F(H)$ to further speculate the behaviors of $h^{0}_v = F(I_{ext})$. 

Strictly speaking, spin and valley cannot be simply equivalent, considering there are some differences between them. Thus, the analogy between valley and spin is just a crude mean to help understanding. However, given that valley and spin also have some similarities in our model, this analogy is still plausible to some extent. At first, our theory is simply built on a two-band Stoner Hamiltonian where valley only serves as a flavor degree of the energy bands. In principle, replacing the valley index with the spin index has no intrinsic influence on our theoretical analysis and the physical picture in Fig.~\ref{FIG1}. In some previous studies in graphene systems, the polarization of spin and valley flavors is often regarded as isospin magnetism as a whole \cite{Zhou3, Zhou}. Secondly, the spin and valley are found to be locked together due to the presence of proximity-induced Ising SOC \cite{Lin}, which also indicates the effect of valley and spin has some equivalence in the experiment.

In the following part, we will explain the influence on $h_{v}^0$ by $j_{ext}$ based on two fashioned theoretical perspectives: Stoner-Wohlfarth model and Rayleigh law.
\\

\textbf{To analyze the function of $h_{v}^{0}=F(I_{ext})$ from Stoner-Wohlfarth model.} The theory of Stoner-Wohlfarth model is based on the coherent rotation of the magnetization in a single-domain particle \cite{Blundell}. This is a simple theoretical model, but it could illustrate the rationality of our approximation to some extent. As shown in Fig.~\ref{FIG6} (a), a spheroidal single-domain particle endowed with uniaxial anisotropy. The magnetization $\mathbf{M}$ is aligned with an angle $\theta$ between the easy axis. The internal energy density is expressed as a function of $\theta$ as \cite{Fiorillo}:
\begin{equation}
	u_{an}(\theta)=K_u \rm{sin}^2(\theta).
	\label{EqB2}
\end{equation}
Here $K_{u}$ is the anisotropy parameter which is related to the magnetocrystalline anisotropy and shape effects. Then we consider the single domain subjected an applied magnetic field $\mathbf{H}$ making the angle $\theta_{H}$ with the easy axis, and the field interaction energy density is \cite{Fiorillo}:
\begin{equation}
	u_{H}(\theta)=-\mu_0 M H \rm{cos}(\theta_H-\theta).
	\label{EqB3}
\end{equation}
Thus, the total Gibbs free energy density is:
\begin{equation}
	g(\theta)=u_{an}(\theta)+u_{H}(\theta)=K_u \mathrm{sin}^2(\theta)-\mu_0 M H \mathrm{cos}(\theta_H-\theta).
	\label{EqB4}
\end{equation}
Note that here we only pay attention on the coherent rotation of $\mathbf{M}$ with the strength of $\mathbf{M}$ unchanged. And the value of $\mathbf{H}$ oscillating between positive and negative values with $\theta_H$ varying between $0^{\circ}$ and $90^{\circ}$. The equilibrium conditions are obtained as $g(\theta)$ reaches the minimum. For convenience, we reduce the $g(\theta)$ as $\tilde{g} (\theta)=g(\theta)/(2K_{u})$. The equilibrium conditions are:
\begin{subequations} \label{AC0}
	\begin{align}
		\frac{d\tilde{g}}{d\theta} &= \frac{1}{2} \mathrm{sin}(2\theta) - h \mathrm{sin}(\theta_H-\theta) = 0 \\
		\frac{d^2\tilde{g}}{d^2\theta} &= \mathrm{cos}(2\theta) + h \mathrm{cos}(\theta_H-\theta) > 0.
\end{align}
\label{EqB5}
\end{subequations}
where $h=\mu_0 M H/2K_{u}=H/H_K$. The solution of Eqs.~(\ref{EqB5}) can be studied analytically in some cases. For example, when $\theta_{H}=0$, the magnetic field $\mathbf{H}$ is aligned with the easy axis, and the solution of Eq.~(\ref{EqB5}a) is $\mathrm{sin}(\theta) = 0$ and $\mathrm{cos}(\theta) = -h$. For the first case, $\theta=0, \pi$ and Eq.~(\ref{EqB5}b) gives $\frac{d^2\tilde{g}}{d^2\theta} = 1 \pm h > 0$. For the second case, $\theta=\mathrm{arccos}(-h)$ and Eq.~(\ref{EqB5}b) gives $\frac{d^2\tilde{g}}{d^2\theta} = h^2-1 >0$. In general, $h < -1, \theta = \pi$;  $h > 1, \theta = 0$; $-1 \leq h \leq 1, \theta = 0$ or $\pi$ (depending on the initial path). To further demonstrate, we plot the magnetization resolved in the field direction $M_{p}(\theta_H)=M \mathrm{cos}(\theta_H-\theta)$ under the cyclic variation of the field $h=H/H_K$ in Fig.~\ref{FIG6}(b) with $\theta_H=0^{\circ}, 15^{\circ}, 45^{\circ}, 75^{\circ}, 90^{\circ}$. The Fig.~\ref{FIG6}(b) is numerically calculated from Eq.~(\ref{EqB5}). It can be found that a change in $h$ causes hysteresis loops where $M_{p}$ can be reversed at certain critical value $h_c$ (i.e. the coercive field). The characteristics of hysteresis loops are strongly dependent on the aligned angle $\theta_{H}$ of $\mathbf{H}$. For $\theta_{H}=0$, we can find a square hysteresis loop with $M_{p}=\pm M$ [see the black solid line in Fig.~\ref{FIG6}(b)]. For the larger $\theta_{H}$, the hysteresis loop shrinks and finally becomes a linear function at $\theta_{H}=90^{\circ}$ [see the magenta solid line in Fig.~\ref{FIG6}(b)]. For an isotropic system of randomly oriented identical particles, the overall mean behaviour stems from an averaged hysteresis loop for different angles.

Next, we refer to the $M_{p}=F(H)$ of the Stoner-Wohlfarth model shown in Fig.~\ref{FIG6}(b), and  analyze the relationship $h^{0}_{v}=F(I_{ext})$. There is a difference between the valley-polarization and the magnetization in ferromagnets. For the latter, a spin-rotation symmetry is maintained and the ferromagnetism is described by a vector order parameter. In contrast, the valley polarization in our system is Ising-like and not a vector \cite{Ying}. The system is either polarized at $K$ valley or $K'$ valley, but never polarized at a valley-coherence state like $\frac{1}{\sqrt{2}}(\vert K \rangle + \vert K' \rangle )$. This means the direction of the valley polarization is only aligned along the easy axis ($z$ axis). In addition, since the applied current $I_{ext}$ will influence the valley polarization but cannot mix two valleys, the effect of $I_{ext}$ should be analogous to the effect of $\mathbf{H}$ at $\theta_{H} = 0^{\circ}$. Therefore, the curve of $h_{v}^{0}(I_{ext})$ should be similar to the curve of $M_p(H)$ at $\theta_{H} = 0^{\circ}$ as shown by black solid lines in Fig.~\ref{FIG6} (b) in a small single-domain valley-polarized system. Actually, even if in a multi-domain system, the averaged hysteresis loop could be somehow like the curve at $\theta_{H} = 0^{\circ}$, because the easy axis of each domain is along the $z$ direction. Therefore, we can conclude that $h_{v}^{0}$ remains nearly unchanged as long as $I_{ext}$ is not too large. Considering the current to flip the valley polarization sometimes demands to reach several tens of nA \cite{Sharpe}, which is basically larger than the critical currents obtained by our numerical calculations, our linear approximation in Eq.~(\ref{Eq5}) has some rationality.
\\
\\
\textbf{To analyze the function of $h_{v}^{0}=F(I_{ext})$ from the Rayleigh law.} For a further comparison, we next refer to another theory called as Rayleigh law, which is used to describe the behavior of ferromagnetic materials at low fields \cite{Bertotti, Dante}. The Rayleigh law is a technical model describing the magnetic hysteresis phenomenon with simple mathematical functions. It quantizes the initial magnetization curve as a second order equation \cite{Rayleigh}:
\begin{equation}
	B(H)=aH+bH^2.
	\label{EqB6}
\end{equation}
Here $a$ corresponds to reversible part of the magnetization process with $a=\lim_{H \to 0} \frac{\partial B}{\partial H}=\mu_0 \mu_i$ ($\mu_i$ is the initial permeability), and $b$ corresponds to the irreversible part of the magnetization process. Based on this initial magnetization curve, Rayleigh law describes the magnetic hysteresis loop by two symmetrical, intersecting parabolic curves \cite{Rayleigh}:
\begin{equation}
	B(H)=(a+bH_{m})H \pm \frac{b}{2}(H^2_{m}-H^2).
	\label{EqB7}
\end{equation}
Note that this function describes the behavior of magnetic induction $B$ with the magnetic field $H$. $H_{m}$ is the amplitude of the scanning magnetic field during the magnetization process. The `+' sign denotes the upper branch of the loop, while the `-' sign denotes the lower branch of the loop. We can draw an analogy from Eqs.~(\ref{EqB6}, \ref{EqB7}), and give the innitial valley polarization curve and the hysteresis for $h_{v}$ as a function of the applied current $j_{ext}$, respectively:
\begin{subequations} \label{AC0}
	\begin{align}
	h_{v}(j_{ext})&=aj_{ext}+bj^2_{ext}\\
	h_{v}(j_{ext})&=(a+bj_{ext,m})j_{ext}\pm \frac{b}{2}(j^{2}_{ext,m}-j^{2}_{ext}).
	\end{align}
\label{EqB8}
\end{subequations}
Similarly, $j_{ext,m}$ is the amplitude of scanning current, i.e. Eq.~(\ref{EqB8}b) is valid when $|j_{ext}| \leq j_{ext,m}$. Once $j_{ext,m}$ is fixed, the form of $h_v(j_{ext})$ is determined by the parameter $a$ and $b$.

In general, the value of $a$ and $b$ can be obtained by experimental fittings. Here, we try to estimate them theoretically. According to Eq.~(\ref{EqB6}), the parameter $a$ reflects the reversible part of the initial magnetization curve, which shows the relationship between $H$ and $M$ as the field strength is increased from a demagnetized magnet ($H=M=0$). To simulate this curve in a valley-polarized system, we use such an expression:
\begin{equation}
	\begin{split}
	h_{v}=\frac{U_{v}}{4\mathcal{V}} \sum_{k, \tau,\tau'} \tau f(\epsilon_{k,\tau}-\tilde{\mu}^{0}-\frac{eV}{2}\mathrm{sgn}(\epsilon'_{k,\tau})-h_{v}^0 \tau \tau').
	\end{split}
\label{EqB9}
\end{equation}
Here $\tau,\tau'=\pm$. Actually, this expression is an average of the initial positive valley polarization $h_v^{0}$ state and initial negative valley polarization $-h_v^{0}$ state, which can be used to simulate a demagnetized state, roughly. When the current is absent ($eV =0$), $h_{v}$ will be zero. In detail, the parameter $a$ is evaluated as:
\begin{equation}
	\begin{split}	
	&a=\left.\frac{\partial  h_{v}}{\partial  I_{ext}}
\right|_{I_{ext}=0}	
=\left.\frac{\partial  h_{v}}{\partial  eV}\right|_{eV=0} \left.\frac{\partial  eV}{\partial  I_{ext}}\right|_{I_{ext}=0}	\\
    &= -\frac{U_{v}}{4\mathcal{V}} \sum_{k, \tau} \tau f(\epsilon_{k,\tau}-\tilde{\mu}^{0}-h_{v}^{0}\tau) \frac{\mathrm{sgn}(\epsilon'_{k,\tau})}{2} \gamma \\
&- \frac{U_{v}}{4\mathcal{V}} \sum_{k, \tau} \tau f(\epsilon_{k,\tau}-\tilde{\mu}^{0}+h_{v}^{0}\tau) \frac{\mathrm{sgn}(\epsilon'_{k,\tau})}{2} \gamma \\
&= -\frac{U_{v}}{4\mathcal{V}} \sum_{k, \tau} \tau f(\epsilon_{k,\tau}-\tilde{\mu}^{0}-h^0_{v}\tau) \frac{\mathrm{sgn}(\epsilon'_{k,\tau})}{2} \gamma \\
&-\frac{U_{v}}{4\mathcal{V}} \sum_{-k, -\tau} (-\tau) f(\epsilon_{-k,-\tau}-\tilde{\mu}^{0} + (-\tau)h_{v}^{0}) \frac{\mathrm{sgn}(\epsilon'_{-k,-\tau})}{2} \gamma \\
&=-\frac{U_{v}}{4\mathcal{V}} \sum_{k, \tau} \tau f(\epsilon_{k,\tau}-\tilde{\mu}^{0}-h^0_{v}\tau) \mathrm{sgn}(\epsilon'_{k,\tau}) \gamma.
	\end{split}
\label{EqB10}
\end{equation}
Here we use the relation: $\epsilon_{k, \tau} = \epsilon_{-k, -\tau}$ and $\epsilon'_{k, \tau} = -\epsilon'_{-k, -\tau}$ . Referring to our derivation of Eq.~(\ref{EqA4}), the parameter $\gamma$ is:
\begin{equation}
\begin{split}
&\gamma = \left.\frac{\partial eV}{\partial j_{ext}} \right|_{j_{ext}=0} \\
&\approx  \frac{h}{e} \sum_{\tau} [-f(\epsilon_{k,\tau}^{max}-\tilde{\mu}^{0}-h_{v}^{0}\tau)+f(\epsilon_{k,\tau}^{min}-\tilde{\mu}^{0}-h_{v}^0\tau)]^{-1}.
\end{split}
\label{EqB11}
\end{equation}

Substituting Eq.~(\ref{EqB11}) into Eq.~(\ref{EqB10}), we can find the value of $a$ is just equal to the value of $\frac{U_v}{2\mathcal{V}}(\alpha_{+}-\alpha_{-})$ as shown in Eq.~(\ref{EqA7}), in view of $\tilde{\mu}=\tilde{\mu}^0$ and $h_v=h^0_v$ at $j_{ext}=0$ ($eV = 0$). For the parameter $b$, it is related to the irreversible part of the initial magnetization curve and cannot be evaluated easily. However, we can assume a case for soft materials where the coercive field $j^{c}_{ext}$ is very small \cite{Blundell}. The coercive field $j^{c}_{ext}$ is the zero point of the function $h_{v}(j_{ext})$, satisfying $(a+bj_{ext,m})j^{c}_{ext} \pm \frac{b}{2}(j^{2}_{ext,m}-(j^{c}_{ext})^2)=0$. We take the case for $`+'$ as an example (the case for $`-'$ is similar) and get:
\begin{equation}
	\begin{split}
j^{c}_{ext}=\frac{(a+bj_{ext,m}) - a\sqrt{1+2\frac{b}{a}j_{ext,m}+2\frac{b^2}{a^2}j^{2}_{ext,m}}}{b}.
\label{EqB12}
	\end{split}
\end{equation}
By using the condition of soft materials ($|j_{ext}^{c}|$ is small), we deduce that $\frac{b}{a}j_{ext,m} \ll 1$ from Eq.~(\ref{EqB12}). Therefore, in the case of soft materials, $a \gg b j_{ext,m}$, Eq.~(\ref{EqB8}b) can be simplified as: $h_{v}(j_{ext})=aj_{ext} \pm \frac{b}{2}j^2_{ext,m} = aj_{ext} \pm h_{v}^{0}$. This just corresponds to the linear relation shown in Eq.~(\ref{Eq5}) in the main text.

Additionally, even if we expand the function in Eq.~(\ref{EqB9}) into the second order of $I_{ext}$, we can find the expansion coefficient:
\begin{equation}
\begin{split}
\tilde{b} &\equiv \left.\frac{\partial^2 h_{v}}{\partial^2 j_{ext}}\right|_{j_{ext}=0} =\left.\frac{\partial^2 h_{v}}{\partial^2 eV}\gamma^2\right|_{eV=0}\\
&=\frac{U_v \gamma^2}{16\mathcal{V}} \sum_{k, \tau} \tau f''(\epsilon_{k,\tau}-\tilde{\mu}^0 -h_{v}^0 \tau)\\ 
&+ \frac{U_v \gamma^2}{16\mathcal{V}} \sum_{k, \tau} \tau f''(\epsilon_{k,\tau}-\tilde{\mu}^0 + h_{v}^0\tau)\\
&=0.
\label{EqB13}
\end{split}
\end{equation}

Although not rigorously, Eq.~(\ref{EqB13}) implies that the coefficient $b$ is small at the bias $eV = 0$. This is some justification to assume $bI_{ext,m} \ll a$ in our case.

In summary, from two fashioned theoretical perspectives, we demonstrate that the linear approximation between $h_{v}$ and $j_{ext}$ in Eq.~(\ref{Eq5}) is still plausible when $j_{ext}$ is relatively small, even though the effect of current or voltage is taken into account in the self-consistent process. Once $j_{ext}$ becomes too large, the weak equilibrium of valley-dependent electron occupations can indeed be broken, and the total valley polarization will be reversed by the flowing current. But in principle, as long as the intersection points shown in Fig.~\ref{FIG3} exist before valley flip happens, our physical picutures are still qualitatively valid.

\section{\label{SEC-C} The effect of trigonal warping effect}
Trigonal warping is a fundamental effect of the energy bands for graphene and (twisted) multilayer graphene systems, which means that the originally rotationally symmetrical Fermi contour (isoenergetical line) is deformed into a shape like the triangle/triangle star, reflecting $C_{3z}$ symmetry of the system \cite{ParkNM,YuanScience,ZhangNai}. In some special cases, the trigonally warped closed Fermi surface may be further broken into three disconnected pockets, which corresponds to a so-called Lifshitz transformation \cite{seilerNature,Zhou}. 

\begin{figure}[ht]
	\includegraphics[width=0.8\columnwidth]{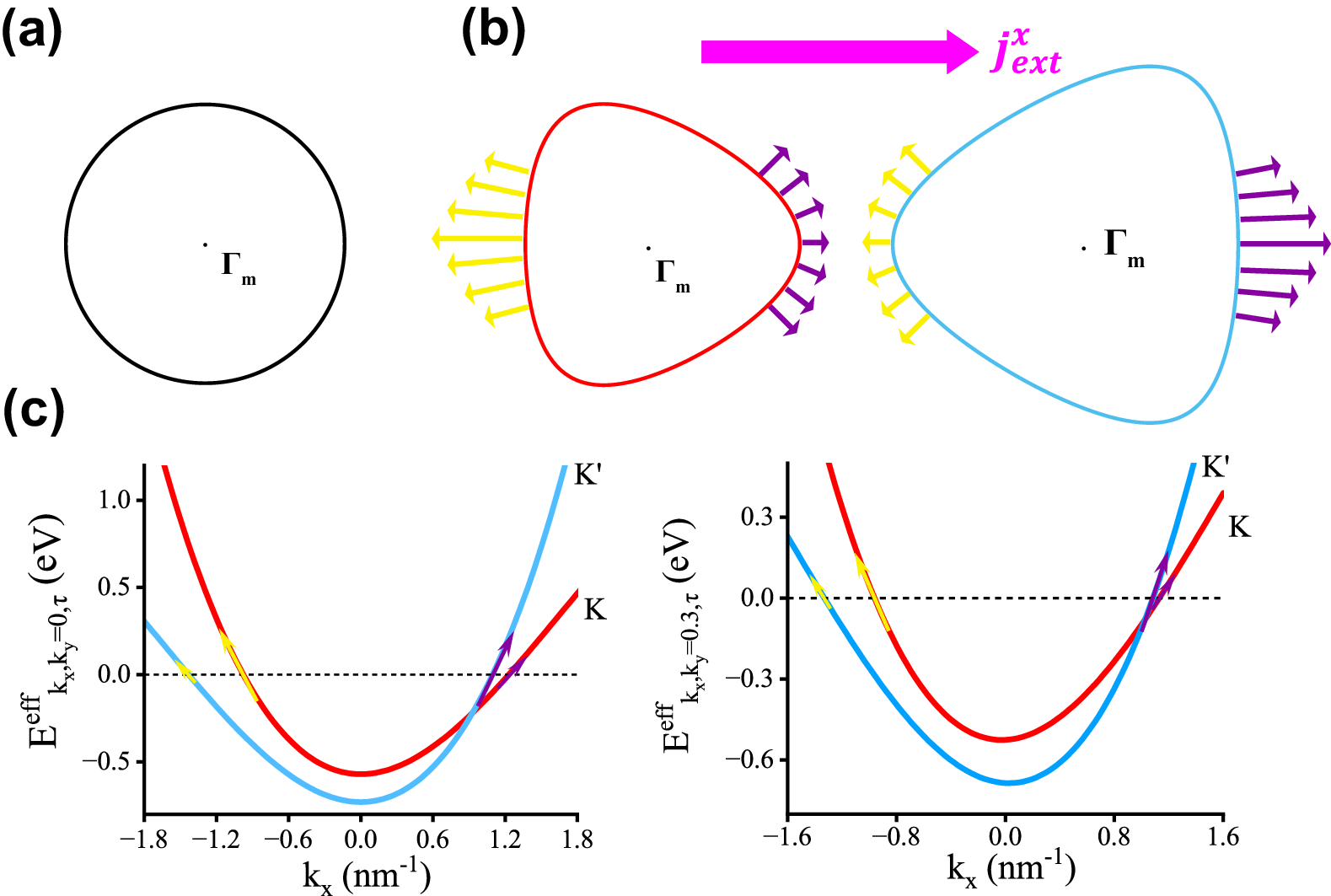}
	\centering %
	\caption{(a) The schematic diagram for an isotropically circular Fermi contour of $E_{\tau}^{eff}$ around $\Gamma_m$ point with $\lambda_1=0$. (b) The schematic diagram for typically trigonally warped Fermi contours around $\Gamma_m$ point of $E_{\tau}^{eff}$ with $\lambda_1 \neq 0$. The red/blue color denotes $K$/$K'$ valley, which are plotted separately for clarity. Purple arrows and yellow arrows schematically indicate local Fermi velocities of right and left movers, respectively. The electric current $j_{ext}^{x}$ is applied along $x$ direction. (c) the effective 1D valley bands $E_{k_x,k_y,\tau}^{eff}$ for the fixed $k_{y}=0$ (left panel) and $k_{y}=0.3$ nm$^{-1}$ (right panel). The colored arrows also schematically indicate amplitudes of local Fermi velocities of effective 1D bands at the Fermi level (dark dashed lines). Here the model parameters (with units) are chosen as $\lambda_0 = 0.5$ eV nm$^{2}$, $\lambda_1=-0.1$ eV nm$^{3}$, $\tilde{\mu}=0.65$ eV, $h_{v}=-0.08$ eV.}
	\label{FIG7}
\end{figure}

In realistic graphene and (twisted) multilayer graphene systems, the trigonal warping effect is one origin of the broken intravalley inversion symmetry within each two-dimensional (2D) valley band. Regardless of whether the Fermi surface has been broken into three pockets, the trigonally warped bands can allow the current-induced valley polarization modulation and the finite-momentum Cooper pairs with a three-fold degeneracy. To see this, we here take a low-energy effective 2D continuum bands $E_{k_x,k_y,\tau}$ near the $\mathbf{\Gamma_m}$ point of the moir$\mathrm{\acute{e}}$ Brillouin zone. A related and more comprehensive tight-binding model will be further presented in Appendix.~\ref{SEC-E}. The discussion is basically similar
for other twisted multilayer graphene systems. The 2D valley bands $E_{k_x,k_y,\tau}$ with a finite valley splitting $h_{v}$ can be written as \cite{Hu,XiePRR}:  
\begin{equation}
E_{k_x,k_y,\tau}^{eff} = \epsilon^{eff}_{k_x,k_y,\tau} -\tau h_v -\tilde{\mu}
=\lambda_0(k_x^2+k_y^2) + \tau \lambda_1 k_x(k_x^2-3k_y^2)-\tau h_v - \tilde{\mu}
\label{EqR1}
\end{equation}
The parameters $\lambda_0$ and $\lambda_1$ denote the kinetic coefficient and trigonal warping coefficient, respectively. $\mathbf{k}=(k_x, k_y)$ is the wave vector relative to the $\mathbf{\Gamma_m}$ point in the $\rm{moir\acute{e}}$ Brillouin zone. Actually, the effective low-energy band $E_{\tau}^{eff}$ can be also rewritten in polar coordinates: $E_{k_r,\phi,\tau}^{eff}=\lambda_0 k_r^2+\lambda_1 k_r^3\cos(3\phi)\tau-\tau h_v-\tilde{\mu}$ with the radial wave vector $k_r = \sqrt{k_x^2+k_y^2}$ and the polar angle $\phi$. We can see the term $\cos(3\phi)$ indeed indicates a three-fold symmetry.

In Figs.~\ref{FIG7}(a,b), we schematically demonstrate two typical types of Fermi contours of $E^{eff}_{k_x,k_y,\tau}$ close to $\Gamma_m$ point with and without the trigonal warping effect. Compared to the isotropically circular Fermi contour ($\lambda_1 = 0$) [Fig.~\ref{FIG7}(a)], the trigonally warped Fermi contour has been deformed into a triangle-like shape with $C_{3z}$ symmetry ($\lambda_1 \neq 0$) [Fig.~\ref{FIG7}(b)]. Additionally, we use red and blue colors to distinguish $K$ band and $K'$ band (plotted separately) in Fig.~\ref{FIG7}(b), and use distinct sizes of Fermi contours to imply a finite valley splitting. When applying an electric current $j_{ext}^{x}$ in the bulk along the $x$ direction, the Fermi level of right movers and left movers should be respectively lifted and declined by the electric voltage. Due to the asymmetry of trigonally warped Fermi contours, we can find the cases of Fermi velocities for right movers (purple arrows) and left movers (yellow arrows) are evidently different, which may also lead to a variation of the carrier occupation within each valley. For convenience, we simply fix the quantum number $k_y$ and regard the 2D effective model as an 1D effective model \cite{Hu}. Due to the symmetry breaking $E_{k_x,\tau}^{eff} \neq E_{-k_x,\tau}^{eff}$, the Fermi velocities along $x$ direction for right movers $v^{+}_{F,\tau}$ and right movers $v^{-}_{F,\tau}$ is usually unequal, which is quite similar to Fig.~\ref{FIG1}(a) in our manuscript. In Fig.~\ref{FIG7}(c), we respectively show the 1D effective bands $E^{eff}_{\tau}(k_x)$ for $k_y=0$ (left panel) and $k_y=0.3$ nm$^{-1}$ (right panel). They all exhibit asymmetrical feature with unequal Fermi velocities (colored arrows) at the Fermi level (dark dashed lines). Due to the time reversal relation between $K$ band and $K'$ band ($E^{eff}_{k_x,\tau}=E^{eff}_{-k_x,-\tau}$), the relative relationship between Fermi velocities of left movers and right movers is also opposite. For example, in Fig.~\ref{FIG7}(c), $v^{+}_{F,\tau} < v^{-}_{F,\tau}$ for $\tau=+$ (red band) while $v^{+}_{F,\tau} > v^{-}_{F,\tau}$ for $\tau=-$ (blue band). This guarantees the opposite modulation of electron occupations induced by the electric currents in two valley bands.

We can also generalize the formulas for current-induced valley polarization modulations in Appendix.~\ref{SEC-A} from 1D model to 2D model. Considering an applied current $j_{ext}^{x}$ along $x$ direction with a bias $V$, the electron occupation $n$ and normal current $j_{ext}^{x}$ can be respectively written as:
\begin{equation}
\begin{cases}
n_{\tau}=\sum_{k_y,k_x} f[E^{eff}_{k_x,k_y,\tau}-\frac{eV}{2}\mathrm{sgn}(\frac{\partial E^{eff}_{\tau}}{\partial k_x})] \\
j_{ext}^{x}=\frac{e}{\hbar \mathcal{V}}\sum_{k_y,k_x,\tau}\frac{\partial E^{eff}_{\tau}}{\partial k_x}f[E^{eff}_{k_x,k_y,\tau}-\frac{eV}{2}\mathrm{sgn}(\frac{\partial E^{eff}_{\tau}}{\partial k_x})]
\label{EqC2}
\end{cases}
\end{equation}
Compared to formulas for 1D model, the formulas for 2D model additionally involves the summation over the quantum number $k_{y}$. Here $\mathcal{V}$ represents the size of the 2D system. Similarly, considering a small bias $V \rightarrow 0$, we can still derive a linear relation between $n_{\tau}$, $h_{v}$ and $j_{ext}^{x}$ (sheet current density), by using a Taylor expansion of $eV$:
\begin{equation}
\begin{cases}
n_{\tau} \approx n_{\tau}^{0}+\alpha_{\tau}^{x}j_{ext}^{x} \\
h_{v} = \frac{U_v}{2\mathcal{V}}(n_{+}-n_{-}) \approx \frac{U_v}{2\mathcal{V}}(\alpha_{+}^{x}-\alpha_{-}^{x})j_{ext}^{x}+h_v^{0}
\end{cases}
\label{EqC3}
\end{equation}
The $n_{\tau}^0$ and $h_{v}^{0}$ denote the initial electron number and valley spitting without the applied current, respectively. The coefficient $\alpha_{\tau}^{x}$ can be arranged as:
\begin{equation}
\alpha^{x}_{\tau}=\frac{\mathcal{V}\hbar}{e}\frac{\sum_{k_x,k_y}f'(E^{eff}_{k_x,k_y,\tau})\mathrm{sgn}(\frac{\partial E^{eff}_{\tau}}{\partial k_x})}{\sum_{k_x,k_y,\tau}\frac{\partial E^{eff}_{\tau}}{\partial k_x}f'(E^{eff}_{k_x,k_y,\tau})\mathrm{sgn}(\frac{\partial E^{eff}_{\tau}}{\partial k_x})}
\label{EqC4}
\end{equation}
For the zero temperature limit and small bias, we can simply approximate the modulation of electron occupation $\Delta n_{\tau}$ for valley $\tau$ as:
\begin{equation}
	\Delta n_{\tau} = n_{\tau}-n_{\tau}^0 \approx \frac{\mathcal{V}}{(2\pi)^2}\frac{eV}{2}\times(\int_{l_1}\frac{dl}{|\mathbf{\nabla}_{\mathbf{k}} E_{\tau}|}-\int_{l_2}\frac{dl}{|\mathbf{\nabla}_{\mathbf{k}} E_{\tau}|})
	\label{EqC5}
\end{equation}
Here $l_{1}$ and $l_{2}$ represent the part of Fermi contour for right movers (purple arrows) and left movers (yellow arrows), respectively. $|\mathbf{\nabla}_{\mathbf{k}} E_{\tau}|$ is related to the magnitude of local Fermi velocity.  Since the trigonal warping breaks intravalley inversion symmetry, the subtraction between two integrals in Eq.~(\ref{EqC5}) is generally nonzero. A finite bias (electric current) can thus induce the modulation of electron occupation in one valley.

\section{\label{SEC-D}The coupling between supercurrents and valley polarizations}

In Fig.~\ref{FIG1}(a) of the main text, we have demonstrated the normal-current-induced valley polarization modulation. The key point is nonequilibrium Fermi levels for moving forward and backward electrons. In this section, we will demonstrate that even an equilibrium supercurrent can also couple to valley polarizations.

In detail, when the system enters the superconducting phase, the supercurrent $j_{s}$ is no longer driven by the finite electric voltage $V$ but instead carried by finite-momentum Cooper pairs. In other words, the coupling between the valley polarization and supercurrents is equal to discuss the influences of the Cooper-pair momentum $2q$ and the superconducting order parameter $\Delta$ on $h_{v}$. To investigate this effect, we simultaneously consider the inter-valley repulsive interaction and inter-valley superconducting pairing in the Hamiltonian, and treat them simultaneously in the mean-field approximation \cite{Banerjee}. Thereby, the total free energy $F_{t}$ should be a combination of free energies shown in Eq.~(\ref{Eq2}) and Eq.~(\ref{Eq9}). It will be a function of order parameters $\Delta(q)$, $h_{v}$ and also the momentum $q$:
\begin{equation}
F_{t}(q,\Delta(q),h_v)=-T\sum_{k,\eta=\pm}\ln(1+e^{-\frac{\tilde{E}_{\eta}(k,q)}{T}})+\sum_{k}E_{-k+q,-}+\frac{\mathcal{V}\Delta(q)^2}{U_s}+\frac{\mathcal{V}h_v^2}{U_v}-\frac{U_v}{\mathcal{V}}n^2+\mu n
\label{EqD1}
\end{equation}
Note that the parameters in Eq.~(\ref{EqD1}) are parallel to those in Eq.~(\ref{Eq2}) and Eq.~(\ref{Eq9}). For simplicity, we set the averaged total electron number $n$ and the chemical potential $\mu$ as constants, which will not influence the total free energy $F_t$.

\begin{figure*}	\includegraphics[width=0.75\textwidth]{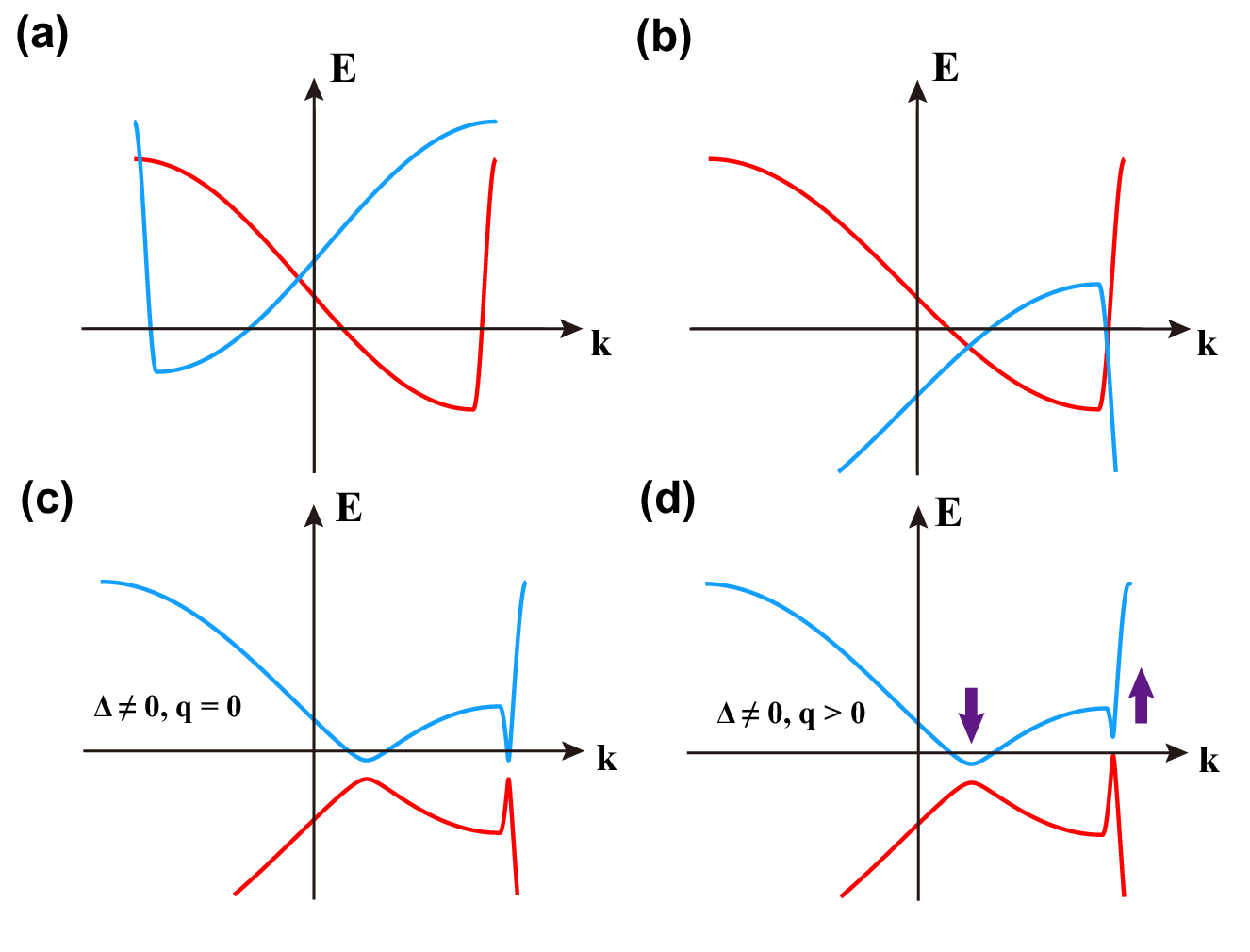}
	\centering
	\caption{
	\baselineskip 14pt
	(a) The schematic diagram for 1D effective valley bands $E_{k,+}$ (red color) and $E_{k,-}$ (blue color) with a finite valley splitting $h_v > 0$. (b) The schematic diagram for $K$ band $E_{k,+}$ (red color) and $K'$ band $-E_{-k,-}$ (blue color) based on the BdG transformation. (c) The schematic diagram for Bogoliubov quasiparticle bands $\tilde{E}_{+}(k)$ (blue color) and $\tilde{E}_{-}(k)$ (red color) with a fixed superconducting order parameter $\Delta \neq 0$. (c) The schematic diagram for Bogoliubov quasiparticle bands $\tilde{E}_{+}(k,q)$ (blue color) and $\tilde{E}_{-}(k,q)$ (red color) with a fixed superconducting order parameter $\Delta \neq 0$ and a finite momentum $q>0$. The purple arrows indicate band shifts, depending on band dispersions of $E_{k,+}$.}
\label{FIG8}
\end{figure*}

To keep the system in the minimum point of the free energy, we demand that the first-order derivatives of $F_{t}$ with respect to $h_{v}$ and $\Delta$ are both zero for the fixed $q$. These lead to a set of self-consistent equations:
\begin{equation}
\left\{
	\begin{array}{lr}  
     \frac{\partial F_{t}(q,\Delta,h_v)}{\partial h_v} = 0, &  \\  
     \frac{\partial F_{t}(q,\Delta,h_v)}{\partial \Delta} = 0, &    
     \end{array} 
\right.
\Rightarrow 
\left\{
	\begin{array}{lr}  
     h_{v}(\Delta,q) = \frac{U_v}{2\mathcal{V}}\sum_{k}\left[f(\tilde{E}_{-}(k,q))-f(-\tilde{E}_{+}(k,q))\right] , &  \\  
     \Delta(q,h_v) = -\frac{U_s}{\mathcal{V}}\sum_{k}\frac{\Delta(q,h_v)}{2\sqrt{E_{2}^2(k,q)+\Delta^2(q,h_v)}}[f(\tilde{E}_{+}(k,q))-f(\tilde{E}_{-}(k,q))]. &    
     \end{array} 
\right.
\label{EqD2}
\end{equation}
Compared to Eq.~(\ref{Eq3}) in the main text, it can be found that the first line of Eq.~(\ref{EqD2}) has corrected the self-consistent expression for $h_{v}$, where the occupation difference between two valley bands $E_{k,\pm}$ is replaced by the occupation difference between two Bogoliubov quasiparticle bands $\tilde{E}_{-}(k,q)$ and $-\tilde{E}_{+}(k,q)$. Once the superconducting order parameter $\Delta$ becomes zero (in the normal phase), the self-consistent equation for $h_{v}$ in Eq.~(\ref{EqD2}) can be verified to be the same as Eq.~(\ref{Eq3}) and the physical picture just returns to Fig.~\ref{FIG1}(a). Especially, once $\Delta$ becomes very large, the Bogoliubov quasiparticle band $\tilde{E}_{-}(k,q)=E_{1}(k,q)-\sqrt{E^2_{2}(k,q)+\Delta^2(q)}$ will be totally negative while $\tilde{E}_{+}(k,q)=E_{1}(k,q)+\sqrt{E_{2}^2(k,q)+\Delta^2(q)}$ will be totally positive. This indicates the summation of $\sum_{k}f(\tilde{E}_{-}(k,q))$ and $-\sum_{k}f(-\tilde{E}_{+}(k,q))$ will be exactly canceled, resulting a zero $h_v$. This phenomenon reflects that the formation of superconducting Cooper pairs will suppress the valley polarization.

Restricting to the single-$q$ order parameter, the supercurrent $J_{s}(q)$ is approximately proportional to $|\Delta|^2 q$ near the superconducting phase transition \cite{Yuan2}. In view of this, to investigate the effect of supercurrents $j_{s}$ on valley polarizations, we consider a non-zero order parameter $\Delta$ and focus on the effect of $q$ on $h_v$. Assuming $q$ is a small quantity, we perform a Taylor expansion for the right part of the first-line equation in Eq.~(\ref{EqD2}):
\begin{equation}
	h_v(\Delta,q)\approx h_v(\Delta,q=0)+\beta(\Delta,q=0)q+O(q^2).
	\label{EqD3}
\end{equation}
Here the first-order expansion coefficient $\beta(\Delta,q=0)$ is derived as:
\begin{equation}
	\begin{split}
\beta(\Delta, q=0)&=\left.\frac{U_v}{2\mathcal{V}}\sum_{k}\left[f'(\tilde{E}_{-}(k,q=0))\frac{\partial \tilde{E}_{-}(k,q)}{\partial q}\right|_{q=0}+\left.f'(\tilde{E}_{+}(k,q=0))\frac{\partial \tilde{E}_{+}(k,q)}{\partial q}\right|_{q=0}\right] \\
&=\frac{U_v}{2\mathcal{V}}\sum_{k}[f'(\tilde{E}_{+}(k,q=0))+f'(\tilde{E}_{-}(k,q=0))]\epsilon'_{k,+}
	\end{split}
	\label{EqD4}
\end{equation}
In Eq.~(\ref{EqD4}), we can find the the first-order expansion coefficient $\beta(\Delta, q=0)$ is still related to the energy band dispersion $\epsilon'_{k,+}$, which is somehow consistent with the linear expansion coefficient $\alpha_{\tau}$ in Eq.~(\ref{EqA7}). Especially, when the valley bands preserve intravalley inversion symmetry: $\epsilon_{k,\tau}=\epsilon_{-k,\tau}$, the summation in Eq.~(\ref{EqD4}) will automatically be canceled, indicating that the finite Cooper-pair momentum $2q$ is uneasy to influence $h_v$. In short, Eq.~(\ref{EqD4}) demonstrates a relation where the valley polarization is approximately proportional to the finite Cooper-pair momentum $2q$ and also the corresponding supercurrent $j_{s} \propto q$.

\begin{figure*}	\includegraphics[width=1\textwidth]{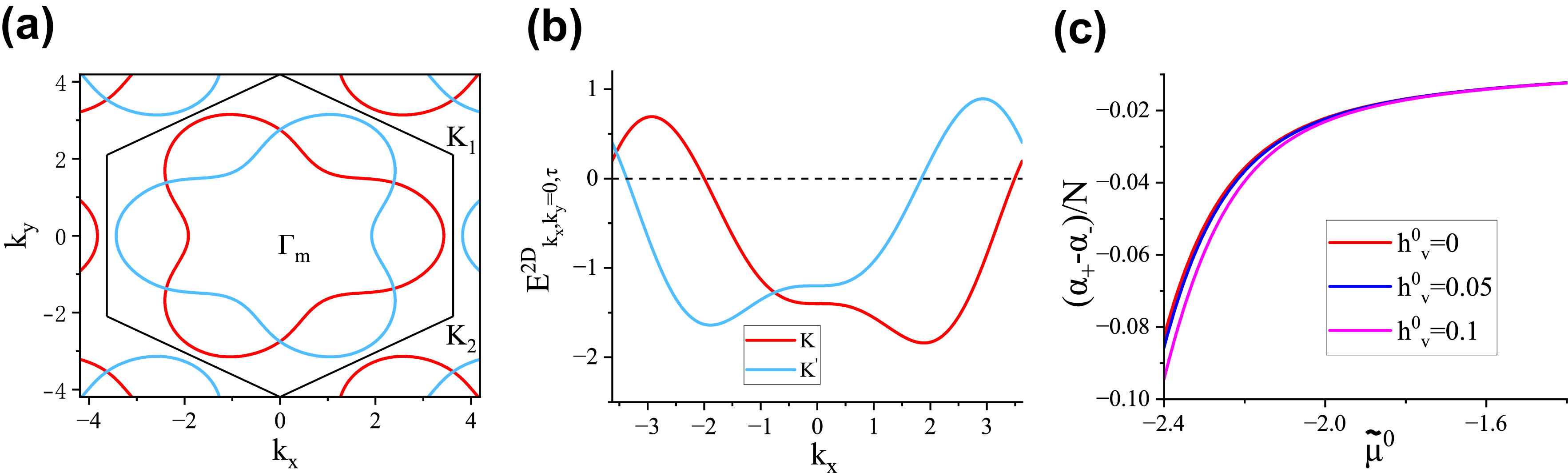}
	\centering
	\caption{(a) The trigonally-warped Fermi surfaces of $H^{2D}_{\tau}(k_x,k_y)$ for effective twisted bilayer graphene bands with $t_{1}=1, t_{2}=0.05, t'_{2}=0.2$, $h_v=0$ and $\mu=-1.4$. The red and blue Fermi surfaces denote $K$ and $K'$ valley, respectively. The hexagonal frame mark the moir$\acute{\textrm{e}}$ Brillouin zone. Note that $k_x$ and $k_y$ are in the units of $L^{-1}_{M}$. Here $L_{M}$ is the moir$\acute{\textrm{e}}$ lattice constant and be set to $L_{M}=1$ in calculations. (b) The 1D effective valley bands of $E^{2D}_{k_x,k_y=0,\tau}$ of $H^{2D}_{\tau}$ with a finite valley splitting field $h_{v}=0.1$. (c) The change of modulation coefficient $(\alpha_{+}-\alpha_{-})/N$ versus the initial chemical potential $\tilde{\mu}^0$ for several initial valley splitting field $h^0_{v}$. Here the coefficient $\alpha_{\tau}$ is in the unit of $h/et_1$.}
\label{FIG9}
\end{figure*}

The similarity between the finite-momentum-induced valley polarization modulation in Eq.~(\ref{EqD2}) and the voltage-induced valley polarization modulation in Eq.~(\ref{EqA6}) can be also elucidated from the BdG Hamiltonian. Performing the Taylor expansion of $q$, the origin BdG Hamiltonian $H(q)$ can be rewritten as:
\begin{equation}
	\begin{split}
H(q)&=\sum_{k}(c^{\dagger}_{k+q,+},c_{-k+q,-})
	\begin{pmatrix}
		E_{k+q,+} & -\Delta(q) \\
		-\Delta(q) & -E_{-k+q,-}
	\end{pmatrix}\binom{c_{k+q,+}}{c^{\dagger}_{-k+q,-}}\\
	&=\sum_{k}(c^{\dagger}_{k+q,+},c_{-k+q,-}),
	\begin{pmatrix}
		\epsilon_{k+q,+}-\tilde{\mu}-h_v & -\Delta(q) \\
		-\Delta(q) & -\epsilon_{k-q,+}+\tilde{\mu}-h_v
	\end{pmatrix}\binom{c_{k+q,+}}{c^{\dagger}_{-k+q,-}}\\
	&\approx \sum_{k,\tau}(c^{\dagger}_{k+q,+},c_{-k+q,-}),
	\begin{pmatrix}
		\epsilon_{k,+}+\epsilon'_{k,+}q-\tilde{\mu}-h_v & -\Delta(q) \\
		-\Delta(q) & -\epsilon_{k,+}+\epsilon'_{k,+}q+\tilde{\mu}-h_v
	\end{pmatrix}\binom{c_{k+q,+}}{c^{\dagger}_{-k+q,-}}.
	\end{split}
	\label{EqD5}
\end{equation}
In Eq.~(\ref{EqD5}), although the finite Cooper-pair momentum does not induce a non-equilibrium electron distribution similar to Fig.~\ref{FIG1}(a), it effectively alters the band structure by an energy shift $\epsilon'_{k,+}q$, which still depends on the sign of band dispersions. Taking 1D effective valley bands shown in Fig.~1(a) as an example, in Fig.~\ref{FIG8}, we schematically demonstrate the effect of this energy shift. In Fig.~\ref{FIG8}(a), the $K$ valley band $E_{k,+}=\epsilon_{k,+}-\tilde{\mu}-h_v$ (red color) and $K'$ valley band $E_{k,-}=\epsilon_{k,-}-\tilde{\mu}+h_v$ (blue color) are respectively plotted with a finite valley splitting $h_v>0$. In Fig.~\ref{FIG8}(b), we convert the $K'$ valley band $E_{k,-}$ into $-E_{-k,-}$ corresponding to the BdG transformation in Eq.~(\ref{EqD5}). By introducing a finite superconducting order parameter $\Delta \neq 0$, the superconducting gap will open, and the $K$ and $K'$ valley bands are recombined into two Bogoliubov quasiparticle bands $\tilde{E}_{+}(k)$ (blue color) and $\tilde{E}_{-}(k)$ (red color). According to Eq.~(\ref{EqD2}), the valley polarization is now related to the occupation difference between $\sum_{k}f(\tilde{E}_{-}(k))$ and $\sum_{k}f(-\tilde{E}_{+}(k))$. Especially, considering a contribution of the small momentum $q$, the energy shift $\epsilon'_{k,+}q$ will break the alignment of right and left parts of BdG bands.  See Fig.~\ref{FIG8}(d), a finite momentum $q$ respectively induces downward and upward energy shift (purple arrows) around right and left crossing points between $E_{k,+}$ and $E_{-k,-}$. When valley bands possess intravalley inversion symmetry, the downward and upward energy shifts should be equal and the occupations on each Bogoliubov quasiparticle band $\tilde{E}_{-}(k,q)$ and $-\tilde{E}_{+}(k,q)$ are approximately unchanged, also making valley polarizations still. Conversely, when intravalley inversion symmetry has been broken, the unequal band dispersions indicate opposing energy shifts cannot be equal, and the occupations on each Bogoliubov quasiparticle band $\tilde{E}_{-}(k,q)$ and $-\tilde{E}_{+}(k,q)$ can be changed. Moreover, the variations of occupations on two Bogoliubov quasiparticle bands may also not be offset [e.g., Fig.~\ref{FIG8}(d)], thereby modulating $h_v$ based on Eq.~(\ref{EqD2}).

\begin{figure}[ht]
	\includegraphics[width=0.85\columnwidth]{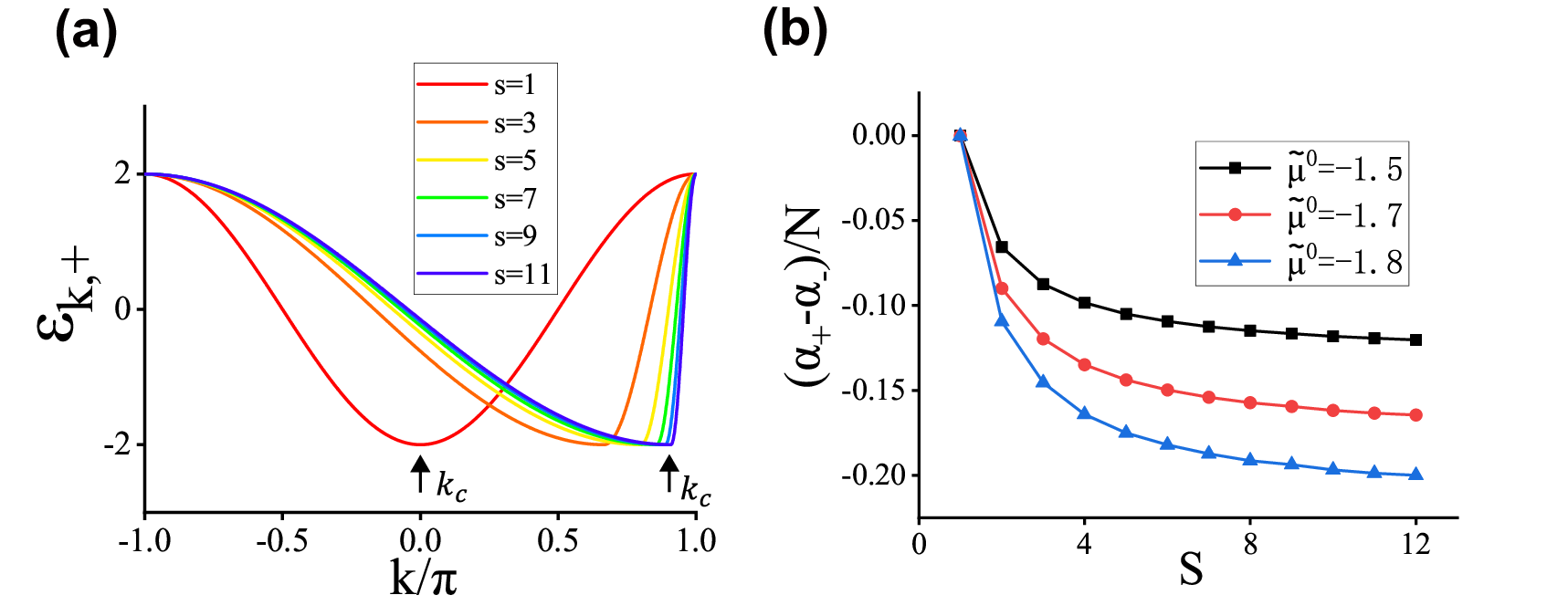}
	\centering %
	\caption{(a) A series of $K$ bands with distinct band asymmetries characterized by $s$. The $K'$ bands are just their TRS counterparts and not shown here. $k_{c}$ denotes the position of the local minimum for the energy band. (b) the change of coefficient $(\alpha_{+}-\alpha_{-})/N$ as a function of $s$ for different $\tilde{\mu}^0$ with $h_{v}^{0}=0.1$.}
	\label{FIG10}
\end{figure}

\section{\label{SEC-E}The estimation of $\alpha_{\pm}$ in a more realistic tight-binding model and the effect of band asymmetry}

In numerical calculations of the main text, we pick an 1D effective toy model to generally demonstrate the broken intravalley inverison symmetry will lead to current-induced valley polarization modulations. Actually, in some more realistic system, the parameters for $\alpha_{\pm}$ in Eq.~(\ref{EqA6}) are likewise significant. Here we use a tight-binding Hamiltonian on the honeycomb lattice with ($p_x, p_y$) orbitals which are often used to simulate the four lowest moir$\acute{\textrm{e}}$ bands of twisted bilayer graphene \cite{Yuan3,koshino,Hu}. It is written as:
\begin{equation}
H^{2D}_{\tau}=\sum_{\left\langle ij\right\rangle}t_1c^{\dagger}_{i,\tau}c_{j,\tau}+\sum_{\left\langle ij\right\rangle'}(t_2-i \tau t'_{2})c^{\dagger}_{i,\tau}c_{j,\tau}+\mathrm{H.c.}-\sum_{i,\tau} (\tilde{\mu}+\tau h_{v}) c^{\dagger}_{i,\tau} c_{i,\tau}.
\label{EqE1}
\end{equation}
where $\left\langle ij\right\rangle$ denotes the nearest-neighbor hopping terms with the hopping energy $t_{1}$, $\left\langle ij\right\rangle'$ denotes the fifth nearest-neighbor hopping terms with the hopping energy $t_2$ and $t'_{2}$. $c^{\dagger}_{i,\tau}$ and $c_{i,\tau}$ respectively denote the creation and annihilation operator for the electron at the lattice site $i$ with $p_{x}+i\tau p_y$ orbital ($\tau=\pm$ represent $K$ and $K'$ valley). Actually, the low-energy effective 2D continuum valley bands in Appendix.~\ref{SEC-C} are just derived from an expansion of $H^{2D}_{\tau}$ at $\Gamma_m$ point where $\lambda_0$ is directly related to $t_1$ and $t_2$, $\lambda_1$ is directly related to $t'_{2}$ \cite{Yuan3, XiePRR, Hu}. Similar to the procedure in our main text, we choose $t_{1}$ as the energy unit ($t_1=1$), and set $t_{2}=0.05 t_1$ and $t'_{2}=0.2 t_1$ according to the previous Reference \cite{Hu}. Usually, for magic-angle twisted bilayer graphene, the hopping energy of $t_{1}$ roughly corresponds to 4 meV \cite{Hu}. Note that $t'_{2}$ characterizes the trigonal warping effect, as is shown by the triangular-shaped Fermi surfaces of $H^{2D}_{\tau}(k_x,k_y)$ in Fig.~\ref{FIG9}(a). It is evident for twisted bilayer graphene bands to break the intravalley inversion symmetry.

For simplicity, we also fix $k_{y}=0$ and reduce the 2D energy bands of Ttwisted bilayer graphene as an 1D energy band. In Fig.~\ref{FIG9}(b), we plot two valley bands $E^{2D}_{k_x,k_y=0,\tau}$ near the Fermi surface with a initial finite valley splitting $h_{v}$. Similar to Fig.~\ref{FIG1}(a), we can see they both exhibit an evident band asymmetry: $E^{2D}_{k_x,k_y=0,\tau} \neq E^{2D}_{-k_x,k_y=0,\tau}$. Analogous to Fig.~\ref{FIG1}(c), we use Eq.~(\ref{EqA7}) to calculate the modulation coefficient $(\alpha_{+}-\alpha_{-})/N$ of the 1D energy band $E^{2D}_{k_x,k_y=0,\tau}$ versus the initial chemical potential $\tilde{\mu}^0$ for several initial valley splitting field $h^0_{v}$. Here $T=0.1t_1$ and $N=2000$ corresponds to the number of discrete $k_{x}$ points. It can be found that $(\alpha_{+}-\alpha_{-})/N$ has demonstrated relatively large values at some chemical potentials.

\begin{figure}[ht]
	\includegraphics[width=0.9\columnwidth]{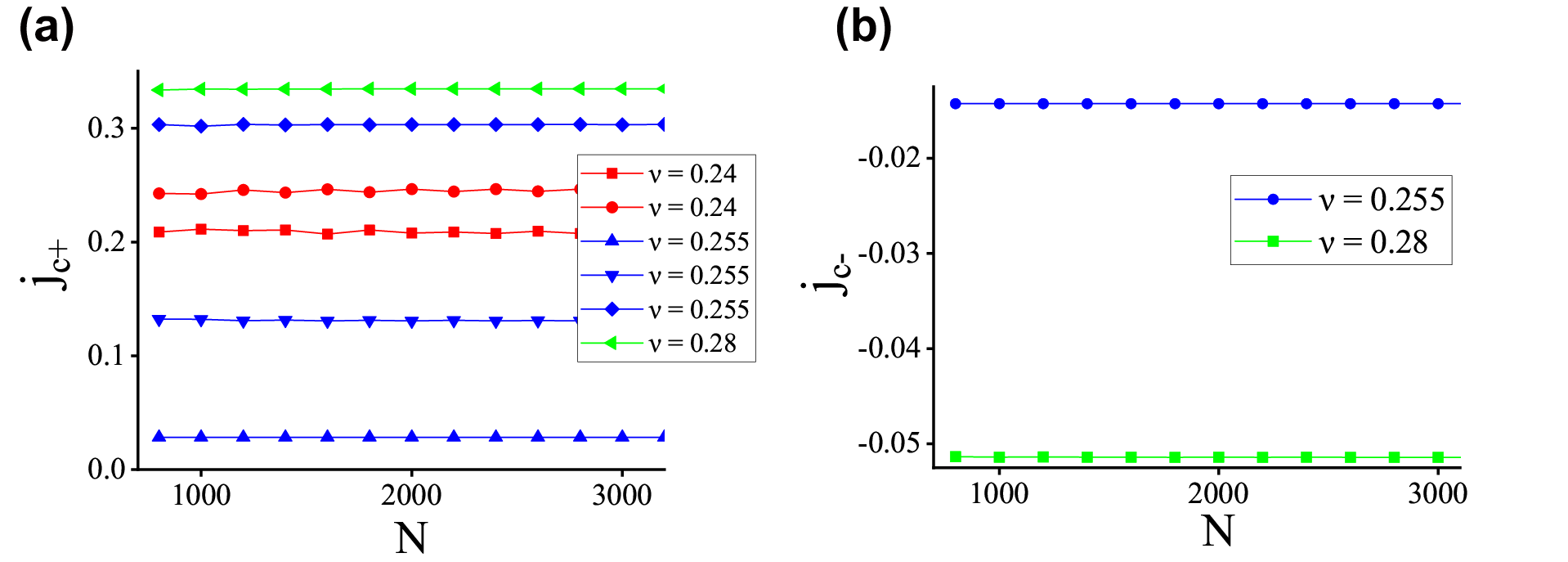}
	\centering %
	\caption{(a,b) the change of actual critical currents $j_{c+}$ (a) and $j_{c-}$ (b) as a function of the system size $N$ for the fixed proportion of the number of electrons $\upsilon=0.24,0.255,0.28$. }
	\label{FIG11}
\end{figure}

In addition, in Fig.~\ref{FIG10}, we also investigate how the band asymmetry affects $\alpha_{+}-\alpha_{-}$. In Fig.~\ref{FIG10}(a), a series of 1D $K$ valley bands are considered: $\epsilon_{k,+}=-2t\cos[\frac{s}{2s-1}(k-\frac{s-1}{s}\pi)]$ for $-\pi \leq k \leq \frac{(s-1)}{s}\pi$ and $\epsilon_{k,+}=-2t\cos(sk-\pi)\times(-1)^{s}$ for $\frac{(s-1)}{s}\pi < k < \pi$. Note that $\epsilon_{k,-}=\epsilon_{-k,+}$.  Here $s$ is introduced to denote the location of the wavevector for the global minimum $k_{c}=\frac{s-1}{s}\pi$ [see Fig.~\ref{FIG10}(a)]. As $s$ increases from 1, $k_{c}$ tends to be close to $\pi$ and the energy band $\epsilon_{k,\tau}$ becomes more asymmetric. For the calculations in the main text, $s$ is set as $s=8$.  In Fig.~\ref{FIG10}(b), under an fixed initial valley splitting field $h_{v}^{0}=0.1$, the magnitude of $(\alpha_{+}-\alpha_{-})/N$ shows an apparent tendency to grow as $s$ climbs, see Fig.~\ref{FIG10}(b) for three different $\tilde{\mu}$. Since a larger coefficient $\alpha_{+}-\alpha_{-}$ implies the current $j_{ext}$ can weaken the valley polarization $h_{v}$ faster, more asymmetric energy bands are more likely to induce the extreme nonreciprocity.

\section{\label{SEC-F}The convergence of results for the system size}
In principle, as long as the proportion of the number of electrons (the filling factor) $\upsilon=n/N$ in the system is fixed, our conclusions in the main text should remain unchanged as $N\to \infty$. To confirm our calculations have converged, we increase the system size $Na$ ($a=1$) by fixing $\upsilon=0.24,0.255,0.28$ respectively. The changes of actual critical current $j_{c+}$ and $j_{c-}$ as a function of $N$ are shown in Fig.~\ref{FIG11}, respectively. Actually, $\upsilon=0.24$ corresponds to $n=480$ when $N=2000$ [Fig.~\ref{FIG3}(d)] where the system enters an extreme nonreciprocity only with two positive $j_{c+}$ [red lines in Fig.~\ref{FIG11}(a)]. $\upsilon=0.255$ corresponds to $n=510$ with $N=2000$ [Fig.~\ref{FIG3}(c)] and the system enters the re-entrant superconductivity with four distinct critical currents $j_{c}$ [dark blue lines in Figs.~\ref{FIG11}(a,b)]. $\upsilon=0.28$ corresponds to $n=560$ with $N=2000$ [Fig.~\ref{FIG3}(b)] where the system enters the conventional SDE with $j_{c+} > 0$ and $j_{c-} < 0$ [light green lines in Fig.~\ref{FIG11}]. Fig.~\ref{FIG11} clearly indicate actual critical currents $j_{c}$ remain nearly unchanged as the system size $N$ varies from 800 to 3200. 

\end{appendix}

%%%%%%%%% END TODO: CONTENTS

%%%%%%%%%% TODO: BIBLIOGRAPHY
% Provide your bibliography here. You have two options:

%%% FIRST OPTION
% Write your entries here directly, following the example below, including:
% Author(s), Title, Journal Ref. with year in parentheses at the end, followed by the DOI number.

%\begin{thebibliography}{99}
%\bibitem{1931_Bethe_ZP_71} H. A. Bethe, {\it Zur Theorie der Metalle. i. Eigenwerte und Eigenfunktionen der linearen Atomkette}, Zeit. f{\"u}r Phys. {\bf 71}, 205 (1931), \doi{10.1007\%2FBF01341708}.
%\bibitem{arXiv:1108.2700} P. Ginsparg, {\it It was twenty years ago today... }, \url{http://arxiv.org/abs/1108.2700}.
%\end{thebibliography}

%%% SECOND OPTION
% Use your bibtex library, formatted by the SciPost style file.
\bibliography{ref.bib}

%%%%%%%%%% END TODO: BIBLIOGRAPHY

\end{document}